\documentclass[twocolumn]{aastex63}
\usepackage{amssymb,amsmath}
\usepackage{graphicx}
\usepackage{xcolor,natbib}
\usepackage{multirow}
\usepackage[T1]{fontenc}
\usepackage{ae,aecompl}
\usepackage{newtxtext, newtxmath}
\usepackage{float}
\usepackage{subfigure}
\usepackage{longtable}
\usepackage{enumitem}
\usepackage{longtable,listings}
\usepackage{longtable}
\usepackage{booktabs}
\usepackage{subfigure}
\usepackage{multirow}
\usepackage[flushleft]{threeparttable}
\usepackage{parcolumns}
\usepackage{rotating}
\def\oiii{[O~{\sc iii}]}
\def\nii{[N~{\sc ii}]}

\submitjournal{ApJ}

\shorttitle{velocity offset between absorption and narrow emission lines}
\shortauthors{Zheng et al.}

\begin{document}
	
\title{Dual core system candidates: a sample of objects with large velocity offset between absorption and narrow emission lines}

\correspondingauthor{Xueguang Zhang; Qirong Yuan}
\email{xgzhang@gxu.edu.cn; yuanqirong@njnu.edu.cn}

\author{Qi Zheng}
\affiliation{School of Physics and Technology, Nanjing Normal University, No. 1,	Wenyuan Road, Nanjing, 210023, P. R. China}

\author{Xueguang Zhang$^{*}$}
\affiliation{Guangxi Key Laboratory for Relativistic Astrophysics, School of Physical Science and Technology,
	GuangXi University, No. 100, Daxue Road, Nanning, 530004, P. R. China}

\author{Qirong Yuan$^{*}$}
\affiliation{School of Physics and Technology, Nanjing Normal University, No. 1,	Wenyuan Road, Nanjing, 210023, P. R. China}
\affiliation{University of Chinese Academy of Sciences, Nanjing 211135, China}

\begin{abstract}
We present a sample of 28 objects at $z<0.3$ from Data Release 16 of the Sloan Digital Sky Survey (SDSS DR16) with large velocity offset (> 200 km/s) of narrow H$\beta$ and H$\alpha$ emission lines relative to absorption lines. Diagnostic classification via the Baldwin–Phillips–Terlevich diagram indicates that the sample comprises 12 AGNs, 12 composite galaxies, 3 H{\sc ii} galaxies, and 1 object of uncertain classification. 
A strong correlation is found between stellar mass and velocity dispersion.
We examine the asymmetries of the narrow H$\beta$ and find that the correlation between velocity offset and narrow H$\beta$ skewness is negligible in both blue-shifted and red-shifted systems, suggesting that the rotating disk model may not fully explain the observed kinematics. The sample exhibits an asymmetric velocity offset distribution, with more red-shifted (17) than blue-shifted (11) objects. No significant correlation is observed between velocity offset and line width in blue-shifted systems, while red-shifted systems show a weak anti-correlation for narrow H$\alpha$, which is inconsistent with the outflow model. The similarity in velocity offset between narrow emission lines supports the dual core system. Furthermore, the SDSS photometric images reveal eight objects with two cores and two with merger features. 
Based on the narrow emission line properties, the objects in our sample represent strong candidates for dual core systems exhibiting velocity offset. Extending this property to higher-redshift populations in the near future may facilitate the identification of merging supermassive black hole pairs at earlier cosmic epochs, providing critical constraints on their formation and evolution.
\end{abstract}

\section{Introduction}
Within the framework of hierarchical galaxy evolution, mergers serve as a key mechanism in the formation of massive galaxies \citep{Be80,Si98,Ba10,Gr17}. These systems grow hierarchically as smaller galaxies merge, resulting in a more massive remnant. Given that most galaxies host central supermassive black holes (SMBHs) \citep{Ko95,Ri98,War21}, such mergers naturally lead to the pairing of these black holes, which gradually inspiral toward the mass centre of new system. During this process, the SMBHs remain in a dual phase, driven by gravitational interaction, with separations from kpc scale to pc scale \citep{Vo16,De19}. 
The system then evolves into a binary black hole (BBH) with sub-pc separation \citep{La20,Ko21}, driven by gravitational wave emission, culminating in a merger that produces a larger SMBH.
The identification and characterization of dual (kpc to pc scale) and binary (sub-pc scale) black hole systems is crucial for understanding the co-evolution of galaxies and their central supermassive black holes. However, direct and confirmed observational evidence for such systems remains scarce \citep{Ko03,Ge07,Ko23,He25}, despite indirect observational hints and theoretical predictions.

During the dual phase, gravitational interactions can displace the two black holes from the original center of their host galaxies, causing them to orbit around the new mass center of the remnant system. 
At this stage, the separation between the black holes can be large enough for each to retain an independent narrow-line region (NLR). As the black holes orbit, the systemic velocities of their respective NLRs trace the orbital motion of the black holes within the merger remnant. In contrast, the stellar absorption lines reflect the bulk motion of the host galaxy.
If both black holes retain sufficient gas, two distinct sets of narrow emission lines may be observed, each exhibiting spatial and velocity offsets relative to the absorption features \citep{Liu10,Zh25}. Alternatively, if only one black hole is actively accreting material, a single set of offset emission lines may appear relative to the absorption lines \citep{Zh23}.
This velocity offset is more likely detected in unequal-mass interactions, where the more massive black hole exhibits lower orbital velocities, while the less massive black hole moves faster. Although \citet{St16} suggested that the more massive black hole tends to be more active in such systems, their simulations have limited spatial resolution, which constrains their reliability at small separations. In contrast, higher-resolution simulations \citep{Ca15} demonstrate that when the separation between the two black holes reaches several kpc, the less massive black hole can accrete more efficiently and display stronger nuclear activity than the more massive one. Although the simulation sample in \citet{Ca15} is not large enough to yield definitive statistical conclusions, their results suggest that the less massive one could make relative velocity offsets more easily detectable in moderately unequal-mass systems.

Emission characteristics of the structural component
provide important insights into physical processes.
Current researches primarily rely on the detection of double-peaked narrow emission lines in spectra \citep{Zh04,Xu09,Wa09,Sh11,Ge12,Co12,Zh24,Zh25}. However, such profiles of narrow emission lines are not uniquely diagnostic of galaxies during the dual phase (dual core systems), as they can also originate from other physical mechanisms including rotating disks \citep{Sm12} and biconical outflows \citep{Ml15,Ne16,Ru19} in the NLRs of single active galactic nuclei (AGNs).
To discriminate these competing scenarios, extensive multi-wavelength follow-up studies have been implemented, and several double-peaked objects have been conclusively identified as genuine dual core systems \citep{Fu11,Liu13,Co15}.

Dual core systems may give rise either to double-peaked narrow emission lines or to velocity offsets between single-peaked narrow emission lines and stellar absorption lines (hereafter mentioned as `velocity offset objects'). Nevertheless, the population of objects exhibiting velocity offsets has been reported much less frequently compared to those with double-peaked narrow emission lines. Similar to the case of double-peaked narrow lines, such velocity shifts may also be attributable to a variety of physical mechanisms, including rotating disk \citep{Co14} and outflows \citep{Co09,Co13}.



Nevertheless, objects exhibiting velocity offsets between narrow emission lines and absorption lines are compelling candidates for dual core systems. Indeed, several velocity offset objects have been confirmed to host two compact cores, such as NDWFS J143316.48+353259.3 and NDWFS J143317.07+344912.0 reported by \citet{Co13}, and SDSS J155708.82+273518.74 identified by \citet{Zh23}. From the DEEP2 Galaxy Redshift Survey, \citet{Co09} detected 30 velocity offset objects out of 1,881 red galaxies, with the results most consistent with inspiralling SMBHs in merger remnants. Subsequently, \citet{Co14} identified 351 type 2 AGNs with velocity offsets exceeding 50 km/s in Data Release 7 of the Sloan Digital Sky Survey, and estimated that 4–8\% of them likely host dual cores after accounting for projection effects.

Current studies show that most velocity offset objects exhibit relatively small offset (typically < 200 km/s). We therefore aim to systematically search in SDSS DR16 \citep{Ah20} using stricter criteria to construct a galaxy sample exhibiting more substantial velocity offset.
By systematically studying velocity offset objects, this work can help disentangle the competing physical mechanisms that produce single-peaked and offset emission lines.

In this manuscript, the procedure to collect the sample of velocity offset objects is shown in Section 2, the basic physical parameters of this velocity offset sample are calculated and analyzed in Section 3, the possible explanations for the velocity offset in our sample are shown in Section 4, and the summary and conclusions are given in Section 5.
We have adopted the cosmological parameters of $H_{0}=70 {\rm km/s/Mpc}$, $\Omega_{\Lambda}=0.7$ and $\Omega_{\rm m}=0.3$.
\section{Sample selection}
Following our work in \citet{Zh23}, we conduct a targeted search of galaxy spectra from SDSS DR16 \citep{Ah20} to systematically search velocity offset objects. 
Following the SQL (Structured Query Language) query from \citet{Zh23}, we systematically relax certain selection thresholds to construct a more statistically representative parent sample of candidates of velocity offset objects.
The applied SQL query here is as follows:
\begin{lstlisting}
SELECT MJD, PLATE, FIBERID, Z, SNMEDIAN, 
V_off_balmer, V_off_forbidden
FROM SpecObjAll AS S
JOIN GalSpecLine AS L
ON S.specobjid = L.specobjid
WHERE S.class = 'galaxy' and
S.veldisp>60 and S.veldisp<400 and
S.veldisperr>0 and
S.veldisp> 5*S.veldisperr and
S.Z < 0.35 and S.zwarning = 0 and
S.SNMEDIAN > 10 and
((V_off_balmer_err>0 and abs(V_off_balmer)>150)
or (V_off_forbidden_err>0 and abs(V_off_forbidden)
>150)) and L.h_alpha_flux>5*L.h_alpha_flux_err
and L.oiii_5007_flux>5*L.oiii_5007_flux_err and
L.h_beta_flux>5*L.h_beta_flux_err and
L.h_alpha_flux<8*L.h_beta_flux
\end{lstlisting}
As a result, the parent sample with 246 galaxies is obtained. 
It should be noted that we also incorporate flux constraints on the main emission lines to ensure the selection of galaxies with reliably detected narrow emission line features.
Besides, the constraints on stellar velocity dispersion are aimed to obtain reliable absorption lines even after considering the instrumental resolution of SDSS \citep{Gr05}.

\begin{figure*}
\figurenum{1}
\plotone{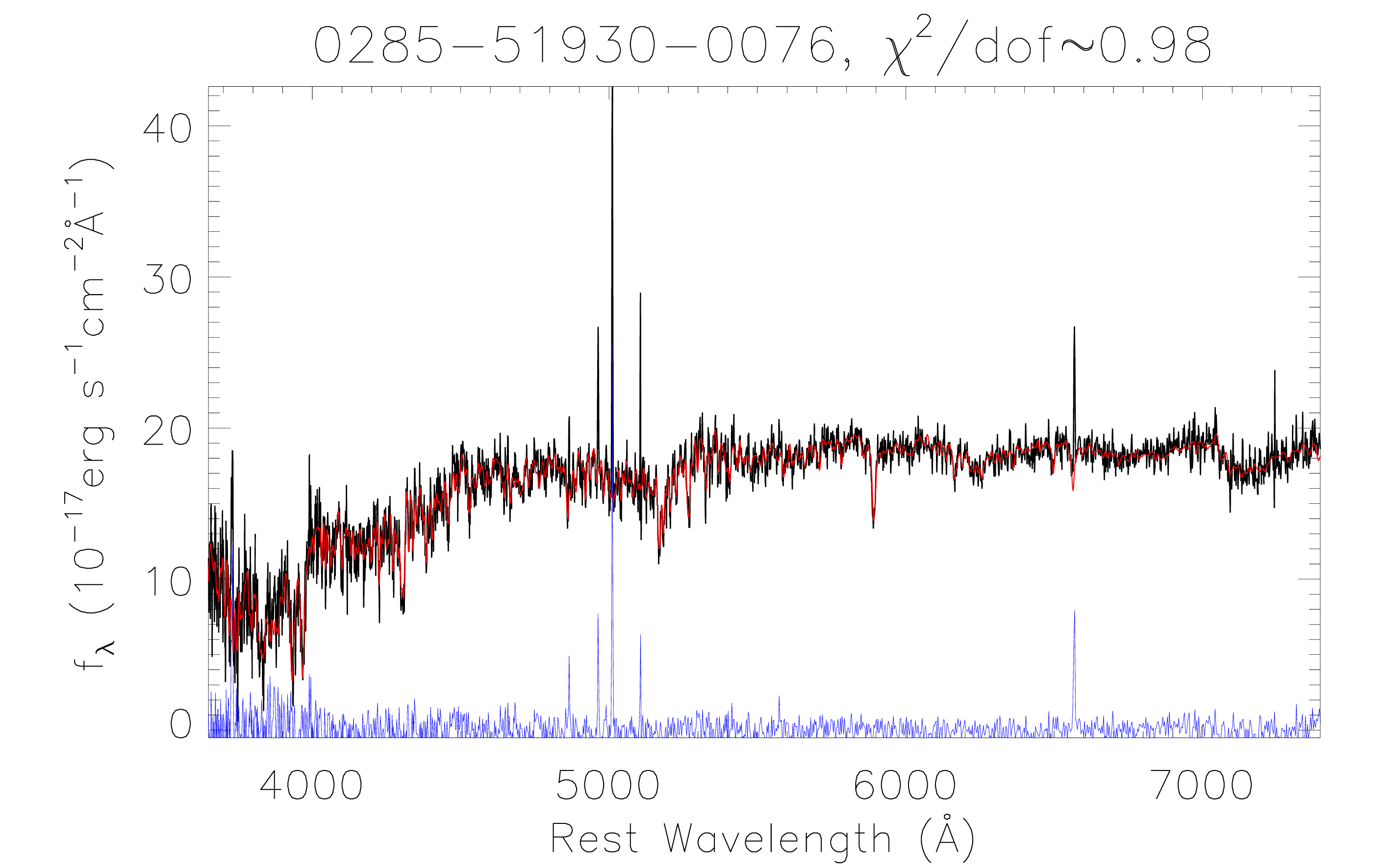}
\caption{An example of the SDSS spectrum in the final sample. The solid black line shows the spectrum with Plate-Mjd-Fiberid shown in the title, the solid red line shows the best fitting results of the starlight determined by the pPXF method with $\chi^2/\rm dof$ shown in the title, and the solid blue line represents the emission-line spectrum calculated by the SDSS spectrum minus the best fitting results. (The complete figure set is available in the online journal.)}\label{fig1}
\end{figure*}

Given the significant stellar continuum contribution from the host galaxy, we employ the penalized pixel fitting (pPXF) method \citep{Me97,Ca04,Ca17} to systematically decompose the observed spectrum into its stellar and non-stellar components.
The pPXF method employs a maximum penalized likelihood approach in pixel space for spectral fitting, with the line-of-sight velocity distributions characterized by Gauss-Hermite parameterization.
More details about the pPXF method can be found in \citet{Ca04,Ca17,Ca23}.
Here, 636 Simple Stellar Population (SSP) templates, consisted of 53 population ages (from 0.03 Gyr to 14 Gyr) and 12 metallicities (from minus 2.27 to 0.4), from the empirical Medium resolution Isaac Newton Library of Empirical Spectra  stellar library \citep{Kn21}, are applied to describe the starlight. Figure \ref{fig1} displays the fitting results for velocity offset objects in the final sample, a subsample selected from the original 246 galaxies.

\begin{figure*}
\figurenum{2}
\plotone{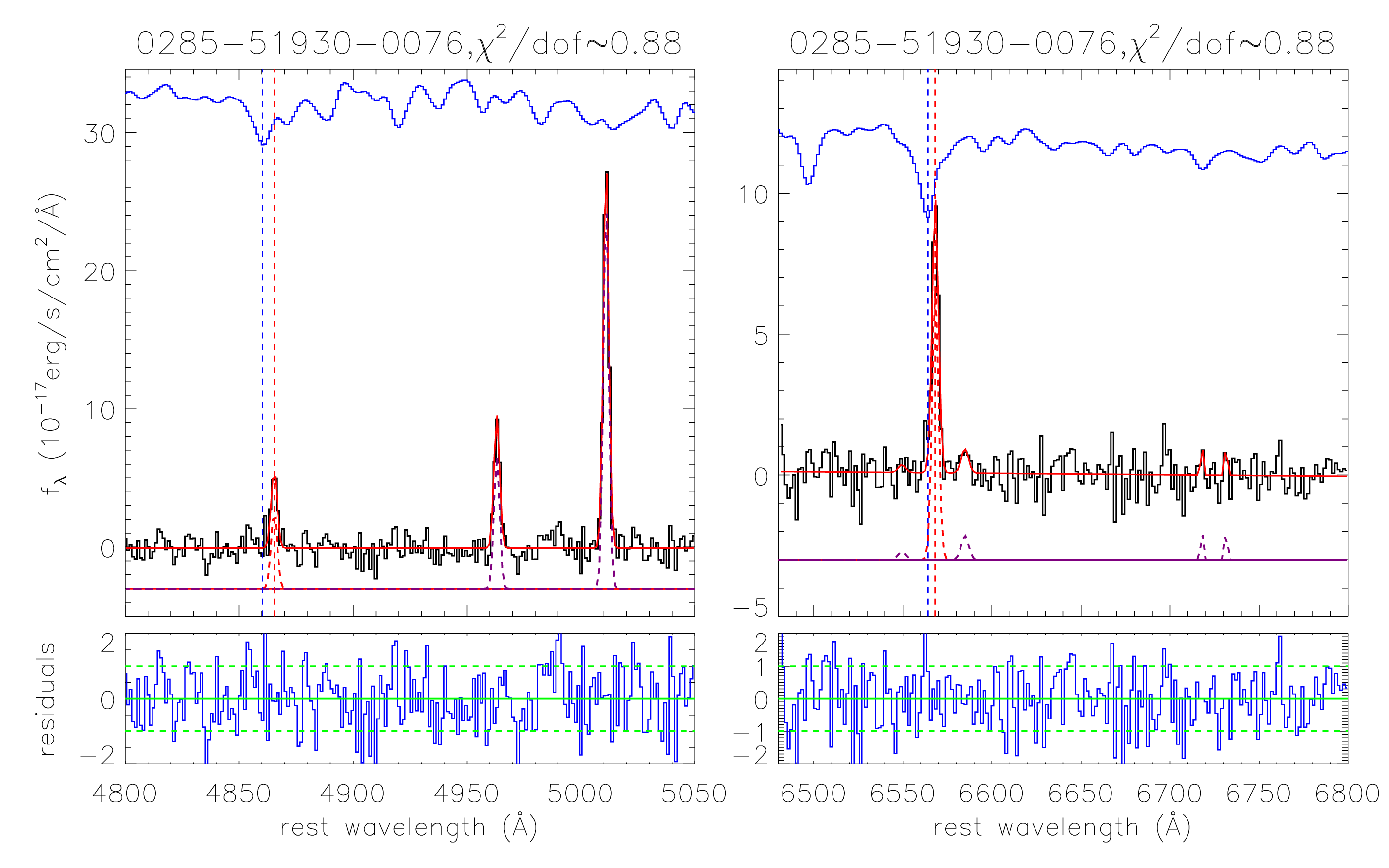}
\caption{An example of the best fitting results of the emission lines around H$\beta$ (left panel) and H$\alpha$ (right panel).
Each of the top panels shows the line spectrum (the solid black line) obtained by subtracting the pPXF determined starlight (the solid blue line) from the SDSS spectrum, the best fitting results of the emission lines as the solid red line with corresponding Plate-MJD-FiberID and $\chi^2/\rm dof$ shown in the title,   
H$\beta$ (H$\alpha$) emission line as the dashed red line, \oiii~([N~{\sc ii}], [S~{\sc ii}]) doublet as the dashed purple lines, and the peaks of absorption and emission lines as the vertical dashed blue and red lines, respectively.
Each of the bottom panels displays the residuals (solid blue line), obtained by subtracting the best-fitting model from the line spectrum and dividing by the SDSS spectral uncertainties. The solid and dashed green lines indicate residual levels of 0 and $\pm$1, respectively. (The complete figure set is available in the online journal.)}\label{fig2}
\end{figure*}

After the subtraction of the starlight, the main emission lines around H$\beta$ (rest wavelength from 4800 to 5050 \AA) and H$\alpha$ (rest wavelength from 6480 to 6800 \AA) are considered.
\oiii, [S{\sc ii}] and [N{\sc ii}] doublets are each modeled with two Gaussian functions, and the two [N{\sc ii}] (\oiii) components are fixed with the same redshift and line width in velocity space, and the flux ratio as 3.
Each of H$\beta$ and H$\alpha$ emission lines is modeled using one Gaussian function.
All fits are performed through the MPFIT package, and the fitting results of the objects in the final sample are shown in Figure \ref{fig2}. 
Given its superior spectral quality, H$\alpha$ absorption line serves as the reference frame for measuring velocity offsets of all emission lines in our analysis.
In final, we retain 28 objects that exhibit robust single-peaked H$\beta$ and H$\alpha$ emission lines (line width and flux three times larger than their uncertainties) and obvious velocity offsets (> 200 km/s) for both narrow H$\beta$ and H$\alpha$ emission lines relative to absorption lines. 
It should be noticed that since the special profiles of \oiii~doublet, the \oiii~flux ratio of the object (Plate-Mjd-Fiberid: 0538-52029-0623) is not fixed.
Furthermore, the [N~{\sc ii}] doublet is not clearly detected in the spectrum of the object (Plate-MJD-Fiberid: 1848-54180-0400).
Due to very broad or weak components around 6730\AA,
[S~{\sc ii}] doublet characterization proves challenging in some objects.
Basic information of the 28 objects is listed in Table \ref{tab1}, and Table \ref{tab2} presents the best fitting parameters (central wavelength, line width, and flux) of narrow H$\beta$ and H$\alpha$ components of the 28 objects .
Meanwhile, the velocity offsets of narrow H$\beta$, H$\alpha$ oiii~emission lines relative to absorption lines ($\upsilon_{\rm em}-\upsilon_{\rm abs}$; $\Delta\upsilon$) are also listed in Table \ref{tab2}.

As shown in Figure \ref{fig2}, the absorption features appear to have broader widths than the narrow emission lines.
This difference can be reasonably explained as discussed below.
The narrow emission lines originate from ionized gas in the NLR ($\sim$1 kpc, \citealt{De22}), where the gas motions are mainly governed by ordered rotation rather than random motions, resulting in potentially small velocity dispersions \citep{Ca17}.
In contrast, the absorption features originate from the integrated stellar light of the host galaxy, and their widths primarily reflect the stellar velocity dispersion ($\sigma_\ast$), which is measured from larger spatial bins than those needed for the more clumpy gas.

A general correlation is found between the stellar velocity dispersion and line width (second moment) of the [O~{\sc iii}] emission line, implying that the gravitational potential of the galactic bulge plays a major role in shaping the gas kinematics in the NLR. However, cases with broader [O~{\sc iii}] lines are also observed, likely due to non-gravitational effects such as outflows \citep{Wo16}. In such cases, the [O~{\sc iii}] profile typically exhibits an asymmetric wing in addition to its narrow core, and a two-component Gaussian model is often adopted to separate these contributions. After correcting for the asymmetric component, the core generally provides a reliable tracer of the stellar velocity dispersion \citep{Gr05}.
Despite this general trend, significant scatter exists in the correlation between the widths of stellar absorption and narrow emission lines. 
We also identify many objects in which the line width (second moment) of narrow Balmer emission lines are noticeably narrower than that of the absorption features, as confirmed by a statistical query of SDSS databases. These findings indicate that while the two quantities are broadly consistent, the deviation from the general trend in some objects can be regarded as reasonable.

Besides, the absorption lines appear not only relatively broad but also asymmetric in Figure \ref{fig2}.
The stellar absorption-line profiles are not always perfectly Gaussian because the line-of-sight velocity distribution (LOSVD) can be influenced by rotation, orbital anisotropy, and stellar population gradients. These effects can produce asymmetric or flat-topped profiles, and in some cases, extended wings. Such non-Gaussian LOSVDs have been extensively discussed in \citet{Ge93, Va93, Ca04}. To account for such deviations, we employ the pPXF method, which models the LOSVD using a Gauss–Hermite expansion. The coefficients $h_3$ and $h_4$ quantify asymmetric and symmetric departures from a Gaussian profile, respectively, and their typical amplitudes are generally below 0.1 \citep{Va93}.

Finally, to ensure the robustness of our stellar continuum fitting and to minimize potential impacts on the measured velocity offsets, we perform an F-test \citep{Ma97, Ge12, Zh25}. The results show that small shift (about 40 km/s) of the template spectrum lead to a noticeable deterioration in the fitting quality, corresponding to higher than 4$\sigma$ confidence level. This indicates that the best-fit parameters provide a reliable description of the velocity shift in the observed stellar spectrum.
Furthermore, we also examine the correlation between our measured stellar velocity dispersions and those provided by the SDSS database. 
As shown in Figure \ref{fig0}, our measurements are consistent with the SDSS database values, with the Spearman rank coefficient of 0.93 and a null-hypothesis probability of $1.97\times10^{-12}$.
Together, these results suggest that the absorption features are well modeled and are unlikely to introduce significant uncertainties in the subsequent calibration of the emission lines or in the measurement of velocity offsets.

\setcounter{figure}{2}
\begin{figure}
\centering\includegraphics[width=8cm,height=8cm]{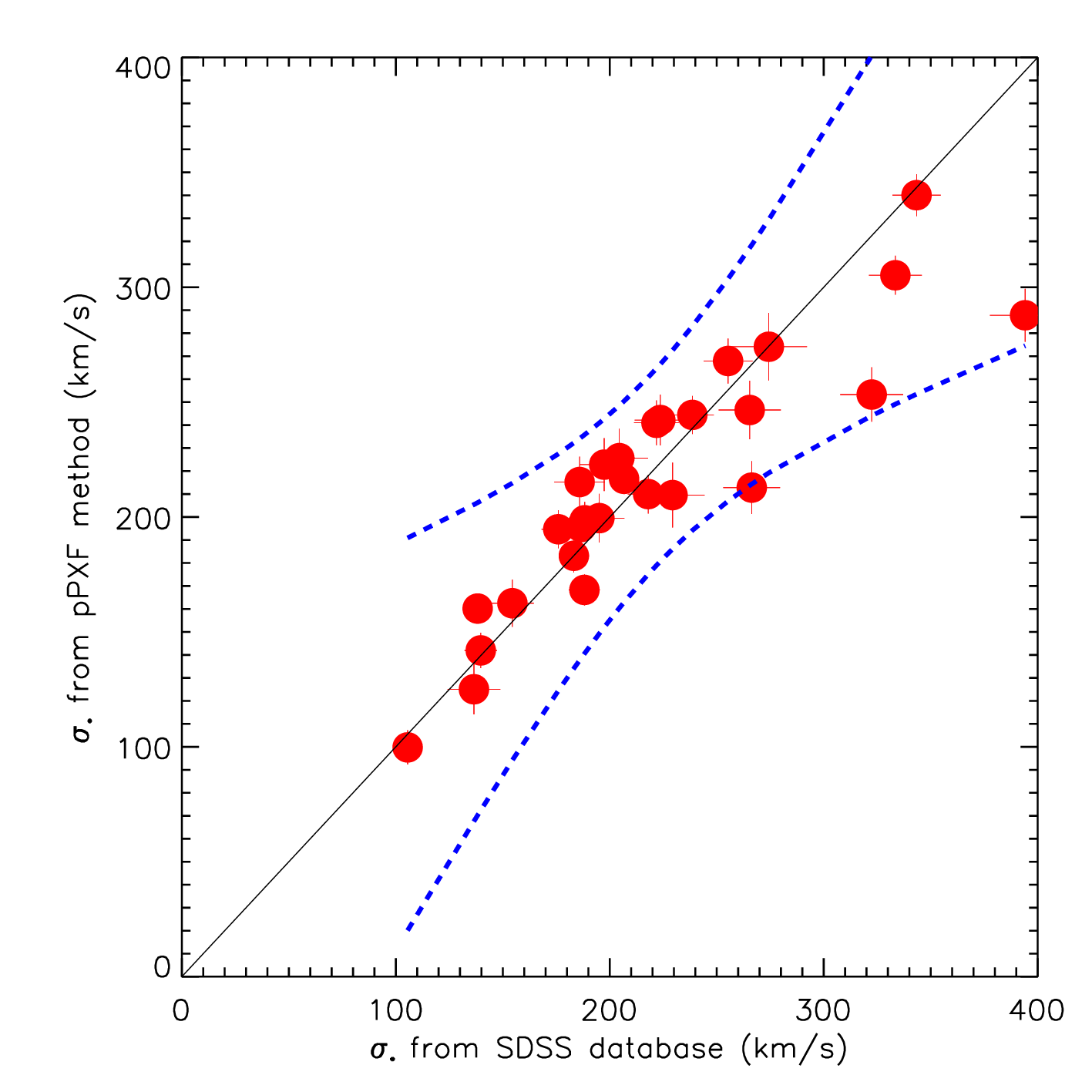}
\caption{The relation between the stellar velocity dispersions of the objects in our sample from the SDSS database and from the pPXF method.
The solid black line indicates the 1:1 relation. The dashed blue lines show 3$\sigma$ confidence bands derived from F-test.
}
\label{fig0}
\end{figure}

\begin{table*}[htbp] 
\centering 
\caption{Basic information} 
\label{tab1} 
\begin{tabular}{cccccccc} 
\toprule
Plate-Mjd-Fiberid & RA & DEC & z& log($\rm M_{\rm BH}$/$\rm M_{\odot}$) & $\sigma_{\ast}$& SDSS $\sigma_{\ast}$ & classification \\ 
(1)&(2)&(3)&(4)&(5)&(6)&(7) &(8)\\ \midrule
0285-51930-0076	&	180.5352245	&	 -0.236922242	&	0.092 	&	7.96 	$\pm$	0.08 	&	141.96 	$\pm$	7.66 	& 139.70 	$\pm$	7.59 	&
	H{\sc ii}	\\ 
0305-51613-0168	&	216.3700498	&	 -0.022942953	&	0.055 	&	7.84 	$\pm$	0.09 	&	168.29 	$\pm$	6.75 	&188.15 	$\pm$	7.31 	&
	Comp	\\ 
0437-51869-0487	&	120.465089	&	43.21235452	&	0.131 	&	7.88 	$\pm$	0.09 	&	209.54 	$\pm$	14.09 	&229.34 	$\pm$	14.88 	&
	Comp	\\ 
0458-51929-0392	&	44.97116769	&	 -7.622101697	&	0.166 	&	8.32 	$\pm$	0.08 	&	246.57 	$\pm$	12.70 	&265.39 	$\pm$	14.49 	&
	AGN	\\ 
0538-52029-0623	&	224.8452057	&	2.01775263	&	0.112 	&	8.15 	$\pm$	0.06 	&	267.90 	$\pm$	9.74 	&255.29 	$\pm$	11.39 	&
	AGN	\\ 
0547-52207-0620	&	126.0068455	&	44.41954303	&	0.129 	&	7.88 	$\pm$	0.09 	&	212.80 	$\pm$	11.41 	&266.36 	$\pm$	13.23 	&
	Comp	\\ 
0569-52264-0022	&	145.0237962	&	3.250073493	&	0.160 	&	8.18 	$\pm$	0.06 	&	222.80 	$\pm$	11.49 	&197.40 	$\pm$	11.53 	&
	AGN	\\ 
0686-52519-0064	&	2.916573175	&	-0.473942329	&	0.058 	&	7.83 	$\pm$	0.09 	&	183.13 	$\pm$	7.28 	&183.21 	$\pm$	6.55 	&
	Comp	\\ 
0742-52263-0540	&	345.7002433	&	15.38539927	&	0.094 	&	8.22 	$\pm$	0.07 	&	244.44 	$\pm$	8.25 	&238.59 	$\pm$	9.96 	&
	AGN	\\ 
0834-52316-0367	&	143.7984065	&	48.01420641	&	0.052 	&	7.88 	$\pm$	0.09 	&	198.58 	$\pm$	7.95 	&188.29 	$\pm$	8.12 	&
	AGN	\\ 
1044-52468-0555	&	212.2827215	&	52.54055703	&	0.114 	&	8.12 	$\pm$	0.06 	&	199.52 	$\pm$	10.56 	&195.12 	$\pm$	11.65 	&
	H{\sc ii}	\\ 
1339-52767-0353	&	247.6232111	&	34.66156199	&	0.136 	&	7.93 	$\pm$	0.08 	&	125.05 	$\pm$	10.91 	&136.51 	$\pm$	12.30 	&
	AGN	\\ 
1341-52786-0061	&	251.6892223	&	31.06208449	&	0.089 	&	8.18 	$\pm$	0.06 	&	195.62 	$\pm$	7.91 	&186.58 	$\pm$	7.63 	&
	Comp	\\ 
1342-52793-0549	&	252.1663693	&	30.01996543	&	0.082 	&	7.76 	$\pm$	0.10 	&	194.68 	$\pm$	8.25 	&175.93 	$\pm$	7.71 	&
	AGN	\\ 
1387-53118-0348	&	229.8781932	&	32.76735095	&	0.115 	&	8.02 	$\pm$	0.07 	&	162.49 	$\pm$	10.27 	&154.48 	$\pm$	10.03 	&
	AGN	\\ 
$\star$1392-52822-0400	&	239.2867697	&	27.58854049	&	0.125 	&	8.95 	$\pm$	0.14 	&	253.30 	$\pm$	11.82 	&322.44 	$\pm$	14.68 	&
	Comp	\\ 
1392-52822-0515	&	240.5331624	&	27.34341664	&	0.150 	&	8.38 	$\pm$	0.08 	&	242.21 	$\pm$	11.04 	&	223.62 	$\pm$	12.00 	&
Comp	\\ 
1463-53063-0137	&	202.3915242	&	45.35147825	&	0.161 	&	8.28 	$\pm$	0.07 	&	215.22 	$\pm$	11.03 	&185.91 	$\pm$	11.76 	&
	Comp	\\ 
1502-53741-0264	&	23.24523263	&	-1.266704312	&	0.096 	&	8.33 	$\pm$	0.08 	&	216.60 	$\pm$	4.47 	&206.67 	$\pm$	4.69 	&
	Comp	\\ 
1669-53433-0541	&	206.9384171	&	51.23446721	&	0.143 	&	8.30 	$\pm$	0.05 	&	241.00 	$\pm$	9.71  	&221.82 	$\pm$	10.45 	&
	Comp	\\
$\ast$1685-53463-0515	&	247.9325305	&	30.80121868	&	0.098 	&	8.09 	$\pm$	0.07 	&	209.97 	$\pm$	8.43 	&217.93 	$\pm$	9.52 	&
	AGN	\\ 
1782-53383-0236	&	124.808619	&	54.5916075	&	0.082 	&	8.03 	$\pm$	0.07 	&	305.20 	$\pm$	8.45 	&333.49 	$\pm$	12.32 	&
	H{\sc ii}	\\ 
1788-54468-0011	&	146.9232891	&	63.66091354	&	0.139 	&	8.26 	$\pm$	0.07 	&	287.80 	$\pm$	11.49 	&394.11 	$\pm$	16.38 	&
	AGN	\\ 
1814-54555-0043	&	223.5856458	&	7.612628726	&	0.130 	&	8.14 	$\pm$	0.06 	&	274.11 	$\pm$	14.69 	&274.35 	$\pm$	17.80 	&
	AGN	\\ 
1848-54180-0400	&	231.6939213	&	27.46797237	&	0.121 	&	8.02 	$\pm$	0.07 	&	225.63 	$\pm$	12.76 	&204.46 	$\pm$	13.37 	&
	None	\\ 
2017-53474-0408	&	205.3173363	&	29.79403832	&	0.116 	&	8.39 	$\pm$	0.08 	&	340.05 	$\pm$	9.12 	&343.38 	$\pm$	11.24 	&
	AGN	\\ 
2104-53852-0491	&	199.6755803	&	31.53746653	&	0.037 	&	7.55 	$\pm$	0.12 	&	160.17 	$\pm$	6.07 	&138.24 	$\pm$	5.90 	&
	Comp	\\ 
2795-54563-0495	&	234.4376126	&	17.24309605	&	0.046 	&	7.38 	$\pm$	0.14 	&	99.76 	$\pm$	7.40 	&105.52 	$\pm$	6.08 	&
	Comp	\\ \hline
\hline
\bottomrule
\end{tabular} \\
Notes. Column 1: Plate, MJD, Fiberid of spectroscopic observation, and the objects with $\star$ and $\ast$ mean that they have been reported in \citet{Zh23} and \citet{Co14}, respectively; Column 2: RA; Column 3: DEC; Column 4: spectroscopic redshift; Column 5: black hole mass; Column 6: stellar velocity dispersion, determined by pPXF method, in units of km/s; 
Column 7: stellar velocity dispersion, collected from SDSS database, in units of km/s; Column 8: classification (H{\sc ii} as H{\sc ii} galaxy, comp as composite galaxy and AGN as active galactic nuclei) of the object determined by the Baldwin-Phillips-Terlevich (BPT) diagram. The object (Plate-Mjd-Fiberid:1848-54180-0400) is not classified due to lacking obvious [N~{\sc ii}].
\end{table*}

\begin{sidewaystable*}[htbp]  
\vspace*{+8cm}
\begin{minipage}{\dimexpr\textheight-8cm\relax} 
\centering 
\caption{Line parameters of H$\beta$ and H$\alpha$ emission lines} \label{tab2} 
\begin{threeparttable}
\begin{tabular}{ccccccccccc} 
\toprule
Plate-Mjd-Fiberid & $\lambda_{H\beta}$  & $\rm \sigma_{H\beta}$  & $F_{H\beta}$ & $\lambda_{H\alpha}$  & $\rm \sigma_{H\alpha}$  & $F_{H\alpha}$ & $\lambda_{abs}$&$\Delta\upsilon_{H\beta}$&$\Delta\upsilon_{H\alpha}$ &$\Delta\upsilon_{\rm[O~{\sc III}]}$\\ 
(1)&(2)&(3)&(4)&(5)&(6)&(7)&(8)&(9)&(10)&(11)\\ 
\midrule
0285-51930-0076	&	4865.4 	$\pm$	0.2 	&	1.3 	$\pm$	0.2 	&	18.4 	$\pm$	2.6 	&	6568.3 	$\pm$	0.1 	&	1.8 	$\pm$	0.1 	&	45.2 	$\pm$	2.5 	&	6563.8 	&	205.1 	$\pm$	19.3 	&	205.6 	$\pm$	11.2 & 206.2 $\pm$	2.7	\\ 
0305-51613-0168	&	4859.5 	$\pm$	0.2 	&	2.6 	$\pm$	0.2 	&	91.5 	$\pm$	5.7 	&	6560.0 	$\pm$	0.1 	&	4.0 	$\pm$	0.1 	&	281.6 	$\pm$	7.7 	&	6566.8 	&	-299.7 	$\pm$	15.9 	&	-313.7 	$\pm$	10.3 & -350.3 $\pm$	7.9	\\ 
0437-51869-0487	&	4864.6 	$\pm$	0.6 	&	2.1 	$\pm$	0.6 	&	11.4 	$\pm$	2.7 	&	6568.6 	$\pm$	0.2 	&	2.4 	$\pm$	0.2 	&	28.0 	$\pm$	2.4 	&	6562.4 	&	220.0 	$\pm$	46.2 	&	286.0 $\pm$ 22.5 & 291.3 $\pm$	22.3	\\ 
0458-51929-0392	&	4865.1 	$\pm$	0.4 	&	1.9 	$\pm$	0.4 	&	21.3 	$\pm$	3.6 	&	6567.3 	$\pm$	0.3 	&	3.8 	$\pm$	0.3 	&	69.4 	$\pm$	5.6 	&	6558.3 	&	438.5 	$\pm$	32.9 	&	412.1 	$\pm$	25.9 &  303.9 $\pm$	32.4	\\ 
0538-52029-0623	&	4864.8 	$\pm$	0.5 	&	2.1 	$\pm$	0.5 	&	16.5 	$\pm$	3.6 	&	6567.3 	$\pm$	0.2 	&	3.1 	$\pm$	0.2 	&	68.9 	$\pm$	4.9 	&	6560.0 	&	341.7 	$\pm$	40.4 	&	333.3 	$\pm$	20.9 & 478.6 $\pm$	24.3	\\ 
0547-52207-0620	&	4858.3 	$\pm$	0.4 	&	2.5 	$\pm$	0.3 	&	18.3 	$\pm$	2.3 	&	6558.6 	$\pm$	0.1 	&	3.3 	$\pm$	0.1 	&	65.0 	$\pm$	2.5 	&	6566.1 	&	-341.5 	$\pm$	31.3 	&	-344.2 	$\pm$	15.9 &  -200.7 $\pm$ 48.0	\\ 
0569-52264-0022	&	4864.1 	$\pm$	0.5 	&	1.7 	$\pm$	0.5 	&	9.6 	$\pm$	2.4 	&	6566.3 	$\pm$	0.2 	&	2.5 	$\pm$	0.2 	&	41.2 	$\pm$	3.4 	&	6558.6 	&	365.7 	$\pm$	41.5 	&	353.9 	$\pm$	20.7 & 327.4 $\pm$	19.7	\\
0686-52519-0064	&	4858.4 	$\pm$	0.1 	&	1.3 	$\pm$	0.1 	&	58.2 	$\pm$	4.3 	&	6558.9 	$\pm$	0.1 	&	1.8 	$\pm$	0.1 	&	234.4 	$\pm$	5.5 	&	6566.4 	&	-343.8 	$\pm$	12.7 	&	-339.4 	$\pm$	8.0 & -208.6 $\pm$	33.5	\\
0742-52263-0540	&	4866.9 	$\pm$	2.0 	&	7.5 	$\pm$	2.0 	&	41.9 	$\pm$	9.8 	&	6565.4 	$\pm$	0.7 	&	6.3 	$\pm$	0.8 	&	90.4 	$\pm$	9.9 	&	6560.0 	&	471.1 	$\pm$	132.8 	&	246.5 	$\pm$	40.0 & 247.3 $\pm$	33.1	\\ 
0834-52316-0367	&	4857.0 	$\pm$	0.3 	&	1.1 	$\pm$	0.3 	&	21.3 	$\pm$	5.0 	&	6557.4 	$\pm$	0.2 	&	1.7 	$\pm$	0.2 	&	76.2 	$\pm$	7.2 	&	6564.7 	&	-354.8 	$\pm$	24.0 	&	-331.9 	$\pm$	13.3 & -351.3 $\pm$	7.2	\\
1044-52468-0555	&	4866.9 	$\pm$	0.3 	&	1.1 	$\pm$	0.3 	&	10.6 	$\pm$	2.6 	&	6569.7 	$\pm$	0.4 	&	3.6 	$\pm$	0.4 	&	59.3 	$\pm$	5.7 	&	6563.8 	&	299.0 	$\pm$	27.8 	&	269.4 	$\pm$	26.1 & 305.1 $\pm$	16.1	\\ 
1339-52767-0353	&	4857.8 	$\pm$	0.2 	&	1.3 	$\pm$	0.2 	&	21.5 	$\pm$	3.2 	&	6558.2 	$\pm$	0.1 	&	1.6 	$\pm$	0.1 	&	56.2 	$\pm$	3.3 	&	6565.8 	&	-356.2 	$\pm$	21.6 	&	-346.6 	$\pm$	12.5 & -340.3 $\pm$	10.0	\\
1341-52786-0061	&	4865.9 	$\pm$	0.3 	&	1.2 	$\pm$	0.3 	&	25.0 	$\pm$	5.3 	&	6568.7 	$\pm$	0.1 	&	1.7 	$\pm$	0.1 	&	118.2 	$\pm$	6.6 	&	6559.4 	&	437.5 	$\pm$	25.9 	&	426.1 	$\pm$	11.7 & 359.8 $\pm$	48.1	\\ 
1342-52793-0549	&	4864.9 	$\pm$	0.3 	&	1.8 	$\pm$	0.3 	&	23.8 	$\pm$	3.5 	&	6567.4 	$\pm$	0.1 	&	2.6 	$\pm$	0.1 	&	100.5 	$\pm$	4.5 	&	6558.3 	&	425.0 	$\pm$	24.5 	&	417.9 	$\pm$	11.7 & 381.5 $\pm$	10.1	\\ 
1387-53118-0348	&	4858.9 	$\pm$	0.3 	&	1.1 	$\pm$	0.3 	&	13.0 	$\pm$	2.7 	&	6560.3 	$\pm$	0.5 	&	2.6 	$\pm$	0.4 	&	33.0 	$\pm$	5.7 	&	6565.5 	&	-273.1 	$\pm$	24.9 	&	-234.5 	$\pm$	29.5 & -189.1 $\pm$	51.3	\\ 
1392-52822-0400	&	4863.2 	$\pm$	0.5 	&	2.9 	$\pm$	0.5 	&	35.8 	$\pm$	5.9 	&	6566.0 	$\pm$	0.2 	&	4.7 	$\pm$	0.2 	&	265.7 	$\pm$	8.4 	&	6556.7 	&	397.4 	$\pm$	43.2 	&	424.9 	$\pm$	18.5 & 384.6 $\pm$	53.9	\\ 
1392-52822-0515	&	4859.6 	$\pm$	0.6 	&	8.2 	$\pm$	0.6 	&	162.5 	$\pm$	11.9 	&	6560.3 	$\pm$	0.3 	&	6.9 	$\pm$	0.3 	&	370.5 	$\pm$	20.3 	&	6565.7 	&	-240.2 	$\pm$	49.1 	&	-249.9 	$\pm$	24.5 & -305.7 $\pm$	21.8	\\ 
1463-53063-0137	&	4864.5 	$\pm$	0.3 	&	1.5 	$\pm$	0.3 	&	16.8 	$\pm$	3.1 	&	6567.1 	$\pm$	0.2 	&	2.0 	$\pm$	0.2 	&	66.8 	$\pm$	4.8 	&	6558.1 	&	406.5 	$\pm$	28.4 	&	410.9 	$\pm$	17.1 & 416.4 $\pm$	4.6	\\ 
1502-53741-0264	&	4858.1 	$\pm$	0.2 	&	1.3 	$\pm$	0.2 	&	34.2 	$\pm$	4.5 	&	6558.5 	$\pm$	0.1 	&	1.8 	$\pm$	0.1 	&	155.1 	$\pm$	5.4 	&	6564.9 	&	-296.4 	$\pm$	15.9 	&	-294.8 	$\pm$	7.3 & -292.4 $\pm$	13.4	\\ 
1669-53433-0541	&	4866.0 	$\pm$	0.8 	&	4.8 	$\pm$	0.9 	&	37.6 	$\pm$	6.1 	&	6565.8 	$\pm$	0.5 	&	6.2 	$\pm$	0.6 	&	84.7 	$\pm$	6.7 	&	6560.1 	&	408.4 	$\pm$	59.6 	&	260.1 	$\pm$	32.1 & 219.5 $\pm$	58.1	\\ 
1685-53463-0515	&	4862.2 	$\pm$	0.9 	&	4.5 	$\pm$	1.0 	&	36.0 	$\pm$	7.1 	&	6565.4 	$\pm$	0.3 	&	6.3 	$\pm$	0.4 	&	179.6 	$\pm$	8.8 	&	6558.7 	&	239.3 	$\pm$	65.3 	&	307.5 	$\pm$	22.1 & 369.2 $\pm$	5.6	\\ 
1782-53383-0236	&	4858.0 	$\pm$	0.1 	&	1.2 	$\pm$	0.1 	&	29.9 	$\pm$	3.1 	&	6558.6 	$\pm$	0.1 	&	2.0 	$\pm$	0.1 	&	112.2 	$\pm$	4.2 	&	6563.4 	&	-233.2 	$\pm$	19.0 	&	-214.9 	$\pm$	14.0 & -198.4 $\pm$	9.0	\\ 
1788-54468-0011	&	4864.2 	$\pm$	0.8 	&	6.0 	$\pm$	0.8 	&	47.3 	$\pm$	6.0 	&	6565.1 	$\pm$	0.3 	&	5.6 	$\pm$	0.3 	&	136.8 	$\pm$	6.1 	&	6557.5 	&	419.3 	$\pm$	60.6 	&	344.8 	$\pm$	23.7 & 383.8 $\pm$	7.9	\\ 
1814-54555-0043	&	4863.1 	$\pm$	0.2 	&	1.0 	$\pm$	0.2 	&	12.1 	$\pm$	2.6 	&	6566.1 	$\pm$	0.3 	&	4.0 	$\pm$	0.3 	&	78.7 	$\pm$	5.5 	&	6559.6 	&	256.1 	$\pm$	28.8 	&	299.0 	$\pm$	27.1 & 74.5 $\pm$	32.6	\\ 
1848-54180-0400	&	4865.9 	$\pm$	0.3 	&	1.2 	$\pm$	0.3 	&	12.6 	$\pm$	2.3 	&	6569.2 	$\pm$	0.2 	&	1.8 	$\pm$	0.2 	&	35.3 	$\pm$	3.4 	&	6558.9 	&	459.4 	$\pm$	26.8 	&	472.6 	$\pm$	20.2 & 469.4 $\pm$	9.5	\\ 
2017-53474-0408	&	4858.1 	$\pm$	0.5 	&	3.1 	$\pm$	0.5 	&	47.0 	$\pm$	6.4 	&	6557.2 	$\pm$	0.3 	&	4.3 	$\pm$	0.2 	&	193.2 	$\pm$	10.4 	&	6567.3 	&	-404.3 	$\pm$	38.5 	&	-462.2 	$\pm$	20.8 & -481.7 $\pm$	9.0	\\ 
2104-53852-0491	&	4864.4 	$\pm$	0.6 	&	3.2 	$\pm$	0.6 	&	41.4 	$\pm$	6.6 	&	6566.6 	$\pm$	0.4 	&	6.7 	$\pm$	0.5 	&	169.3 	$\pm$	10.2 	&	6559.4 	&	346.3 	$\pm$	39.2 	&	328.6 	$\pm$	23.8 & 332.2 $\pm$	32.3	\\ 
2795-54563-0495	&	4858.3 	$\pm$	0.2 	&	1.1 	$\pm$	0.2 	&	17.2 	$\pm$	2.8 	&	6558.4 	$\pm$	0.1 	&	1.6 	$\pm$	0.1 	&	56.5 	$\pm$	3.2 	&	6564.6 	&	-271.1 	$\pm$	17.9 	&	-283.1 	$\pm$	9.7 & -264.6 $\pm$	14.2	\\ \hline
\bottomrule
\end{tabular} 
\begin{tablenotes}
\item Notes. Column 1: Plate, MJD, Fiberid of spectroscopic observation; Column 2 and Column 3: central wavelength and corresponding line width (the second moment) of narrow H$\beta$ in units of \AA; Column 4: corresponding flux of narrow H$\beta$ in units of $10^{-17}{\rm erg/s/cm^{2}}$; Column 5 and Column 6: central wavelength and corresponding line width (the second moment) of narrow H$\alpha$ in units of \AA; Column 7: corresponding flux of narrow H$\alpha$ in units of $10^{-17}{\rm erg/s/cm^{2}}$; Column 8: peak of H$\alpha$ absorption line in units of \AA, Column 9: velocity offset between narrow H$\beta$ emission line and reference frame determined by H$\alpha$ absorption line in units of km/s; Column 10: velocity offset between narrow H$\alpha$ emission line and reference frame determined by H$\alpha$ absorption line in units of km/s.
Column 11: velocity offset between \oiii~and reference frame determined by H$\alpha$ absorption line in units of km/s.
\end{tablenotes}
\end{threeparttable}
\end{minipage}
\end{sidewaystable*}

\section{Basic properties}
In this section, it presents measurements and analysis of fundamental properties of the 28 objects exhibiting significant velocity offset in our sample.
\subsection{velocity offset}
\begin{figure*}
	\centering\includegraphics[width=8cm,height=6cm]{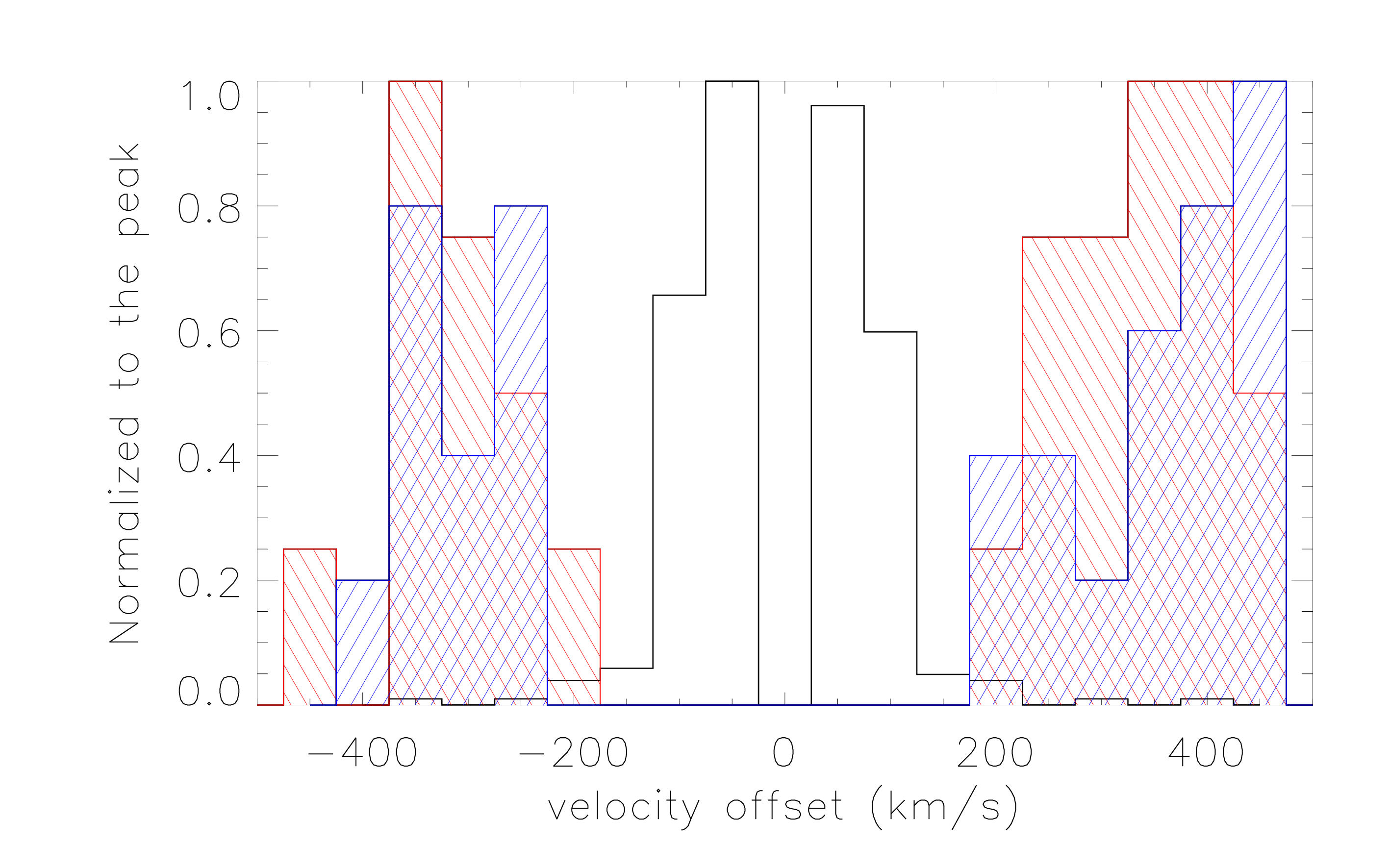}
	\centering\includegraphics[width=8cm,height=6cm]{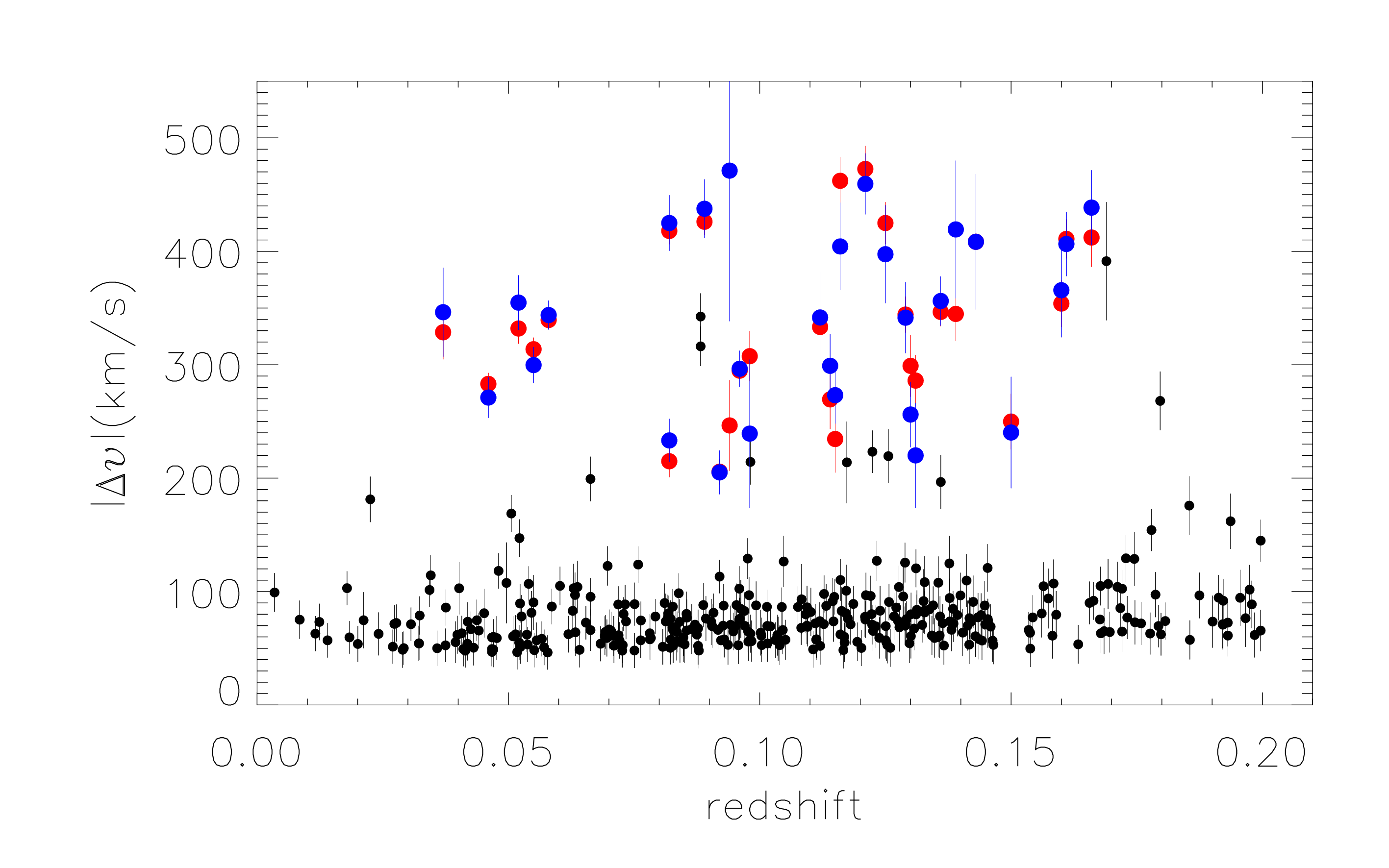}
	\caption{The distributions of velocity offset (left panel) and the relationship between redshift and velocity offset (right panel) in both our sample and \citet{Co14} sample. In the left panel, the solid black lines show the distributions of $\Delta\upsilon$ of \citet{Co14} sample, the histogram filed by blue lines and red lines show the distributions of $\Delta\upsilon$ of narrow H$\beta$ and H$\alpha$ emission lines in our sample, respectively. In the right panel, the solid circles in blue and in red mark the results of narrow H$\beta$ and H$\alpha$ emission lines in our sample, respectively. The solid circles in black mark the results of Balmer emission lines in \citet{Co14} sample.
	}
	\label{fig3}
\end{figure*}

There is a large sample of 351 objects with narrow emission lines showing velocity offset relative to absorption lines \citep{Co14}, which provides an excellent velocity-offset reference, and our work aims to extend this foundation further.
The velocity offset distributions of narrow H$\beta$ and H$\alpha$ in our sample and of Balmer emission lines in \citet{Co14} sample (listed in Table 1) are shown in the left panel of Figure \ref{fig3}.
It is clear that the objects in our sample exhibit significantly larger velocity offsets.

There is a weak correlation between redshift and $\mid$$\upsilon_{\rm em}-\upsilon_{\rm abs}$$\mid$ with the Spearman Rank correlation coefficient as 0.30 and null-hypothesis probability of 20.50\% in narrow H$\beta$ emission lines, and 
with the Spearman Rank correlation coefficient as 0.31 and null-hypothesis probability of 11.32\% in narrow H$\alpha$ emission lines of our sample.
This correlation may arise from observational selection effects: higher-redshift galaxies must be more luminous and have greater mass to be detected, and then the most massive merging systems exhibit larger orbital velocities, thereby producing a positive trend between emission-line velocity offset and redshift.

Moreover, this correlation also can be found in \citet{Co14} sample with the Spearman Rank correlation coefficient as 0.25 and null-hypothesis probability of $10^{-4}\%$.
However, when restricting the analysis to objects with velocity offset less than 200 km/s in the \citet{Co14} sample (quantity as 326), we find only a negligible correlation between velocity offset and redshift (the Spearman Rank correlation coefficient as 0.13, null-hypothesis probability of 0.04\%). This statistically weak trend, contrasting with the stronger dependencies observed in our sample, suggests possible fundamental differences in the nature of velocity offset between the two samples or at least very different sample collection effects in the two samples. Therefore, the subsequent analysis focuses exclusively on the properties of our sample with larger velocity offset than 200 km/s.

A potential concern in fiber spectroscopy is the aperture effect, where the measured velocity from emission or absorption lines could be biased if the fiber does not centrally sample the galaxy or captures only one side of a kinematically resolved region. This could artificially amplify the velocity offset between narrow emission lines and absorption lines.
However, the SDSS fiber placement is highly accurate, and for our sample the fiber centers are consistent with the target coordinates, with no significant deviations (the largest offset only about $0.30^{\prime\prime}$, after checking the SDSS provided parameters of $plug\_ra$ and $plug\_dec$). 
Given the SDSS spatial resolution of $0.396^{\prime\prime}$ pixel, the separation between the two nuclei in a dual core system must be very small to remain unresolved.
Thus, the observation is not strongly biased toward one side, and it is unlikely to capture emission from only one side of NLR or host galaxy.
Our sample is based on SDSS spectra obtained with a fiber of $3^{\prime\prime}$ diameter. At the minimum redshift of our sample ($z\sim0.037$, $1^{\prime\prime}\sim0.73$ kpc), the fiber semidiameter ($1.5^{\prime\prime}$) corresponds to $\sim$1.1 kpc. This scale is comparable to or larger than the typical size of the NLR ($\sim$1 kpc; \citealt{De22}). The fiber therefore captures most NLR emission and reduces aperture bias for the majority of objects.
Besides, as mentioned in Section 4.3, eight objects in our sample show two compact cores, and two show tidal tails. These morphologies indicate strong gravitational interactions where large relative velocities are expected. In addition, the  velocity offsets are rather large, and systematic across the whole sample rather than limited to individual cases, making it unlikely that aperture losses alone can explain the observed trends.	
Moreover, to assess the possible impact of aperture effects, we divide the sample into two subsamples at the median redshift. If aperture effects were significant, higher-redshift objects should exhibit systematically narrower narrow emission lines, since their NLRs are more fully covered by the fiber. However, the Student's t-test technique shows no significant difference in the mean widths of narrow H$\alpha$ and H$\beta$ between the two subsamples. We also compare the mean velocity offsets of H$\alpha$, H$\beta$, and \oiii, and found no substantial differences. These results suggest that aperture effects, if present, have little influence on our conclusions.
While a minor contribution from this effect cannot be excluded in individual cases, it is unlikely to dominate the systematic velocity offsets observed across the whole sample.

\subsection{Baldwin–Phillips–Terlevich diagram}
\begin{figure}
\centering\includegraphics[width = 8cm,height=6cm]{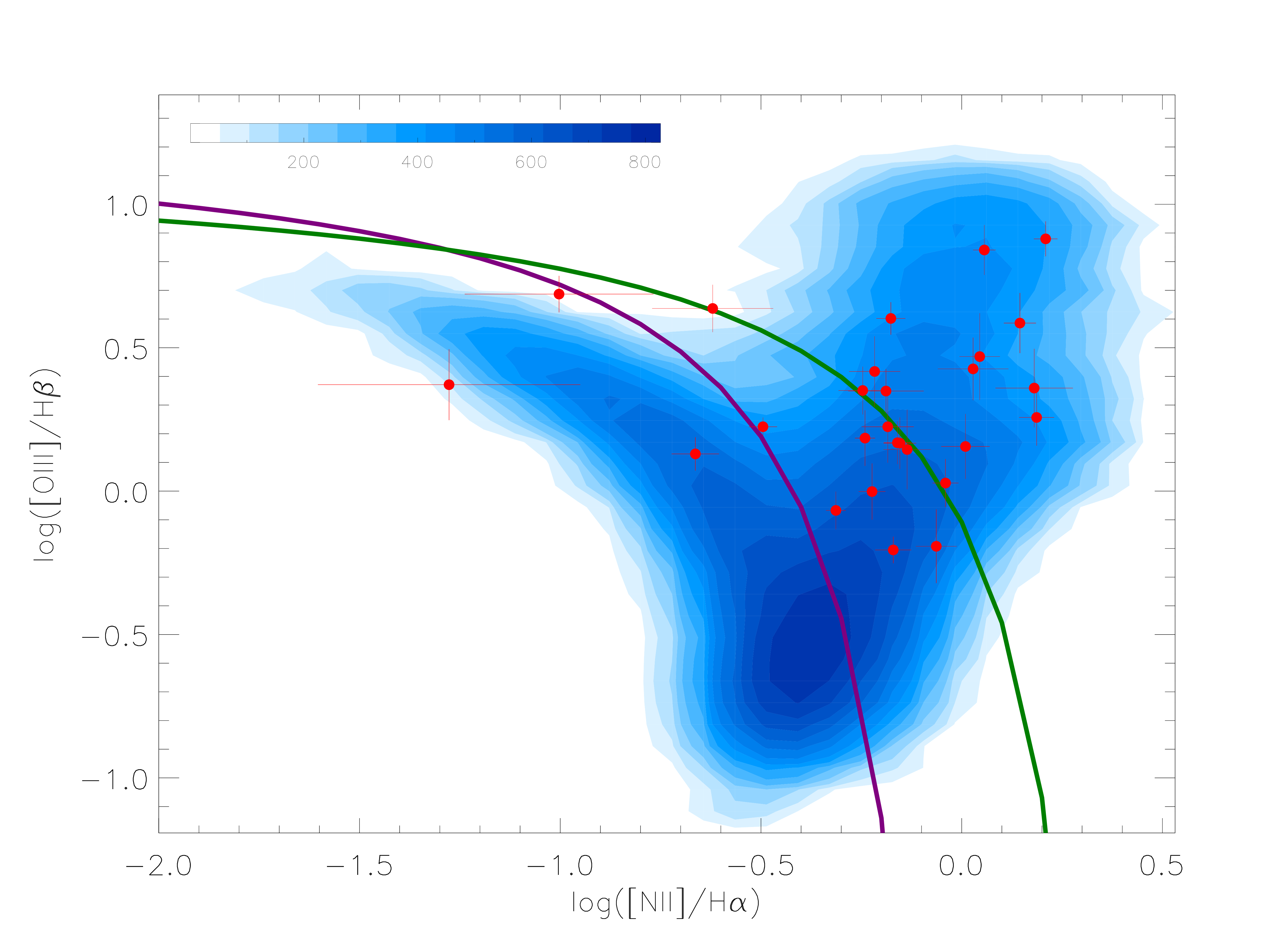}
\caption{The BPT diagram for the 27 objects (in red color) with velocity offset in our sample. 
The blue contour shows the distribution of more than 35,000 narrow emission-line galaxies in SDSS DR15, as shown in \citet{Zh20}.
The solid purple and green lines show the dividing lines reported by \citet{Ka03} and \citet{Ke01}, respectively.
}
\label{fig4}
\end{figure}

We subsequently classify these 28 objects using the Baldwin–Phillips–Terlevich (BPT) according to the diagnostic criteria \citep{Ke01,Ka03,Ch25}, with the classification results presented in Figure \ref{fig4} and Table \ref{tab1}. The [S~{\sc ii}]/H$\alpha$ versus \oiii/H$\beta$ diagram is not employed for this analysis due to the absence of significant [S~{\sc ii}] emission lines in a substantial fraction of the sample.
As a result, there are 12 AGNs, 12 composite galaxies, 3 H{\sc ii} galaxies, and 1 unknown-type objects due to lacking obvious [N~{\sc ii}] emission lines.

\subsection{black hole mass}
Many studies propose that galaxies and their dark matter halos may be fundamentally characterized by either stellar mass or velocity dispersion \citep{Mo11,Li13,Bo15}. While stellar mass and velocity dispersion originate from distinct physical processes, both parameters exhibit fundamental connections to dark matter halo properties.
It is undeniable that there is a strong correlation between stellar mass and velocity dispersion \citep{Za16,Da22}. 
The Spearman Rank correlation coefficient between the total stellar mass and the stellar velocity dispersion determined by the pPXF method in our sample is 0.73, with null-hypothesis probability of $10^{-4}$\%.

In addition, the well-established scaling relations between supermassive black hole masses and host galaxy properties provide compelling evidence for SMBH feedback regulating galaxy evolution \citep{Ko13,Su20}. This connection manifests most prominently in massive galaxies, where central black hole masses correlate tightly with bulge properties \citep{Ma98}, particularly through the well-documented $M_{\rm BH}-M_{\rm \star,bul}$ (black hole mass and bulge stellar mass) relation \citep{Yo15,Se21}.

\begin{figure}
\centering\includegraphics[width =8cm,height=6cm]{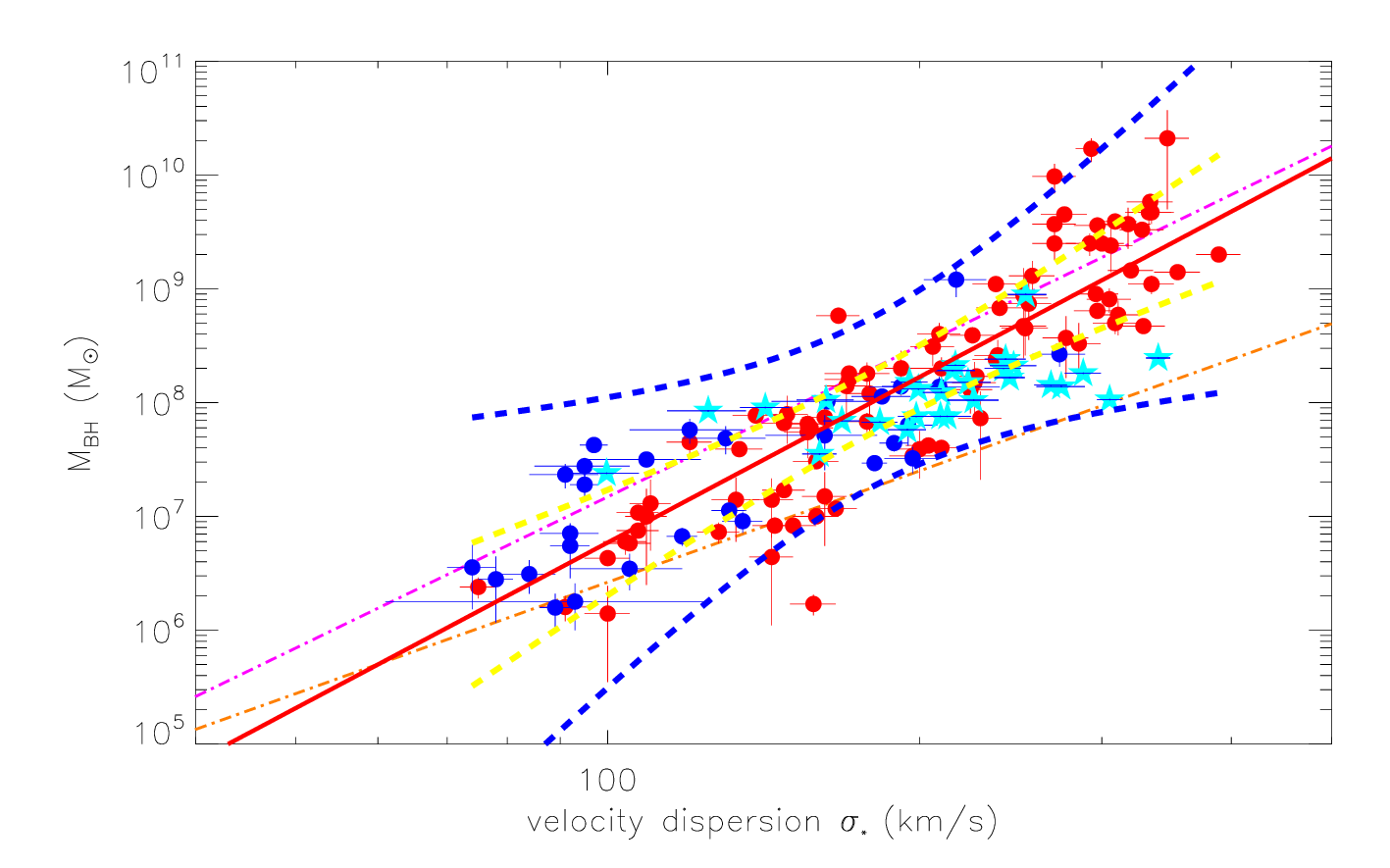}
\caption{$M_{\rm BH}-\sigma_{\ast}$ relation for the 28 objects (five-pointed star in cyan) in our sample. 
The $\sigma_{\ast}$ here for our sample is determined using the pPXF method.
The red and blue solid circles denote the results of 89 quiescent galaxies and 29 reverberation-mapped AGNs, respectively.	
The dot-dashed magenta and orange lines indicate the $M_{\rm BH}-\sigma_{\ast}$ relations from \citet{Ko13} and \citet{Ho14}, respectively. 
The best fitting result of the 89 quiescent galaxies and the 29 reverberation mapped AGNs, as discussed in \citet{Zh23}, is shown as the solid red line, while 3$\sigma$ and 5$\sigma$ confidence bands derived from F-test are displayed as dotted yellow and blue lines, respectively.	
}
\label{fig5}
\end{figure}

If accepted that the total stellar mass is scaled to the bulge mass \citep{Bl14,Re15,De20}, the $M_{\rm BH}$ can be estimated from the $M_{\rm BH}-M_{\rm \star,bul}$ relation in \citet{Sa11}, and the results are shown in Table \ref{tab1}. 
Besides, there is a well-studied relation between stellar velocity dispersion and black hole mass, which is known as $M_{\rm BH}-\sigma_{\ast}$ relation \citep{Fe00,Ge00,Gu09}.
While numerous studies demonstrate that AGNs follow the same $M_{\rm BH}-\sigma_{\ast}$ relation as quiescent galaxies \citep{Gr04,Gr08,Ji22}, other work suggests potential variations across galaxy types \citep{Gr09,Ra17,Do21}. 
Our sample provides an opportunity to examine the $M_{\rm BH}$–$\sigma_\ast$ relation. As shown in Figure \ref{fig5}, the distribution of our objects is broadly consistent with the relation observed for inactive galaxies and AGNs. The measured values lie within the 5$\sigma$ confidence level of these systems, indicating that our sample does not deviate significantly from the established scaling relation.

\subsection{skewness}
To precisely quantify spectroscopic emission-line asymmetry, we use a skewness statistic S based on the emission-line spectrum after subtracting the host galaxy contribution. The emission-line spectrum, represented as a discrete distribution function comprising n data points with flux values ($f_i$) at corresponding pixel coordinates ($x_i$), serves as the basis for this calculation. The S \citep{Ka06} is defined as:
\begin{equation}
\begin{split}	
s=\frac{\sum\limits_{i}^n(x_{i}-\overline{x})^3f_{i}}{\sum_{i}^{n} f_{i}\sigma^{3}},
\end{split}
\end{equation}
where $\overline{x}$ is the average $x_{i}$ and $\sigma$ is the line width.

Here, we use the weighted skewness as the statistical skewness, computed over the wavelength range where the flux exceeds 10\% of the peak flux similar as done in \citet{Co14}, and the weighted narrow H$\beta$ skewness is calculated and shown in the top-left panel of Figure \ref{fig6}.
The skewness uncertainty here is quantified through a bootstrap approach: (1) Randomly selecting 50\% of spectral data points to generate 1000 resampled spectra; (2) Recalculating the weighted skewness for each new spectrum using the same equation; (3) Deriving final uncertainties from the half-width at half-maximum of the resulting skewness distribution, which exhibits Gaussian characteristics.

While the narrow H$\beta$ emission line in our spectra is well-fit by a single Gaussian, the significant skewness values observed in some objects reveal underlying asymmetry in the line wings, which show a feature that may be smoothed over by the Gaussian approximation but remains detectable through higher-order statistical moments.
The narrow H$\beta$ emission line profiles of two objects exhibiting significant skewness ($\mid$S$\mid$ > 0.5) are presented in the top-right panel of Figure \ref{fig6}, demonstrating pronounced asymmetries in their wing components.
Due to blending between \nii~and narrow H$\alpha$ emission lines, which introduces systematic uncertainties in skewness measurements, we restrict our analysis to narrow H$\beta$ emission lines.

\section{Discussion}
This section synthesizes the observed properties of velocity offset objects to evaluate potential physical origins. We systematically analyze three plausible scenarios capable of producing such kinematic signatures: rotating disk, outflows and dual core system effects. 

\subsection{rotating disk}
In a rotating disk, a symmetric distribution of Doppler velocities can be expected in the case of relatively low rotational speeds. However, if only part of the disk is observed due to obscuration by dust or the intrinsically clumpy nature of the gas, this symmetry is broken. 
The integrated line profile then becomes both shifted and asymmetric, with its centroid displaced toward the velocity of the visible side (e.g., blue-shifted if the approaching side dominates). The skewness arises because the emission from the unobscured side is both stronger and spans more extreme velocities. Thus, preferential observation of the approaching side produces a blue-shifted line with stronger flux on the blue side (negative skewness), while visibility of the receding side leads to a red-shifted line with positive skewness \citep{Co14}.

In order to test whether such rotating disk is a significant cause of velocity offset in our sample, we use the skewness as measured and shown in Section 3.4 and Figure \ref{fig6}.
For the blue-shifted systems, we find a negligible correlation between $\upsilon_{\rm em}-\upsilon_{\rm abs}$ and narrow H$\beta$ skewness with the Spearman Rank correlation coefficient as 0.12 and null-hypothesis probability of 70.9\%. 
For the red-shifted systems, we also find a negligible correlation between $\upsilon_{\rm em}-\upsilon_{\rm abs}$ and narrow H$\beta$ skewness with the Spearman Rank correlation coefficient as 0.15 and null-hypothesis probability of 56.0\%.
To verify that our results are not biased by the definition of weighted skewness, we recompute the weighted skewness using different flux thresholds (5\% and 20\% of peak flux) to ensure the null correlation with velocity offset is not an artifact of our initial 10\% threshold selection. The consistent results across definitions demonstrate that our sample is not dominated by rotating disk.

\subsection{outflows}
The observed velocity offset in narrow emission lines may alternatively be attributed to a powerful wind within the inner NLR \citep{Ko08}.
Distinct from stellar winds, galactic winds experience increasing mass loading during outward propagation, resulting in progressive deceleration with radial distance from the galactic center.
It reveals stratified NLRs \citep{Da20,Ko22}, 
where emission lines originating nearer to the central region display greater velocity offset while emission lines at further region preferentially show lower velocity offset \citep{Co09}.
Due to obscuration from both the dust torus \citep{Ne15} and the larger-scale dusty interstellar medium \citep{Bo20} of the host galaxy, outflow kinematics manifests predominantly blue-shifted emission lines, as the red-shifted emission lines from the far side are obscured.
However, our sample exhibits an opposite asymmetric distribution of velocity offset, with 17 objects showing red-shifted emission lines compared to only 11 with blue-shift.

\begin{figure*}
\centering\includegraphics[width=18cm,height=9cm]{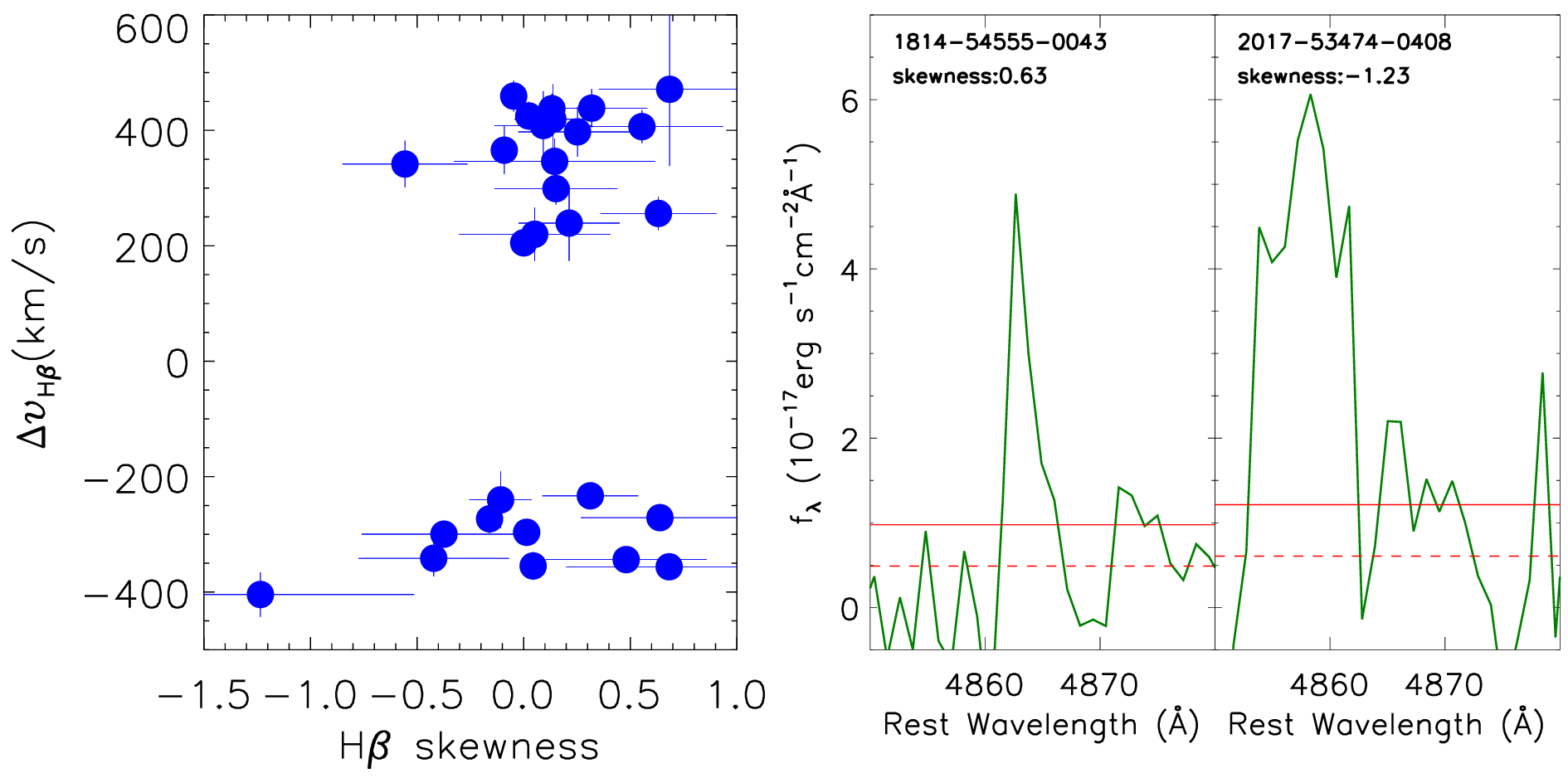}
\centering\includegraphics[width=18cm,height=9cm]{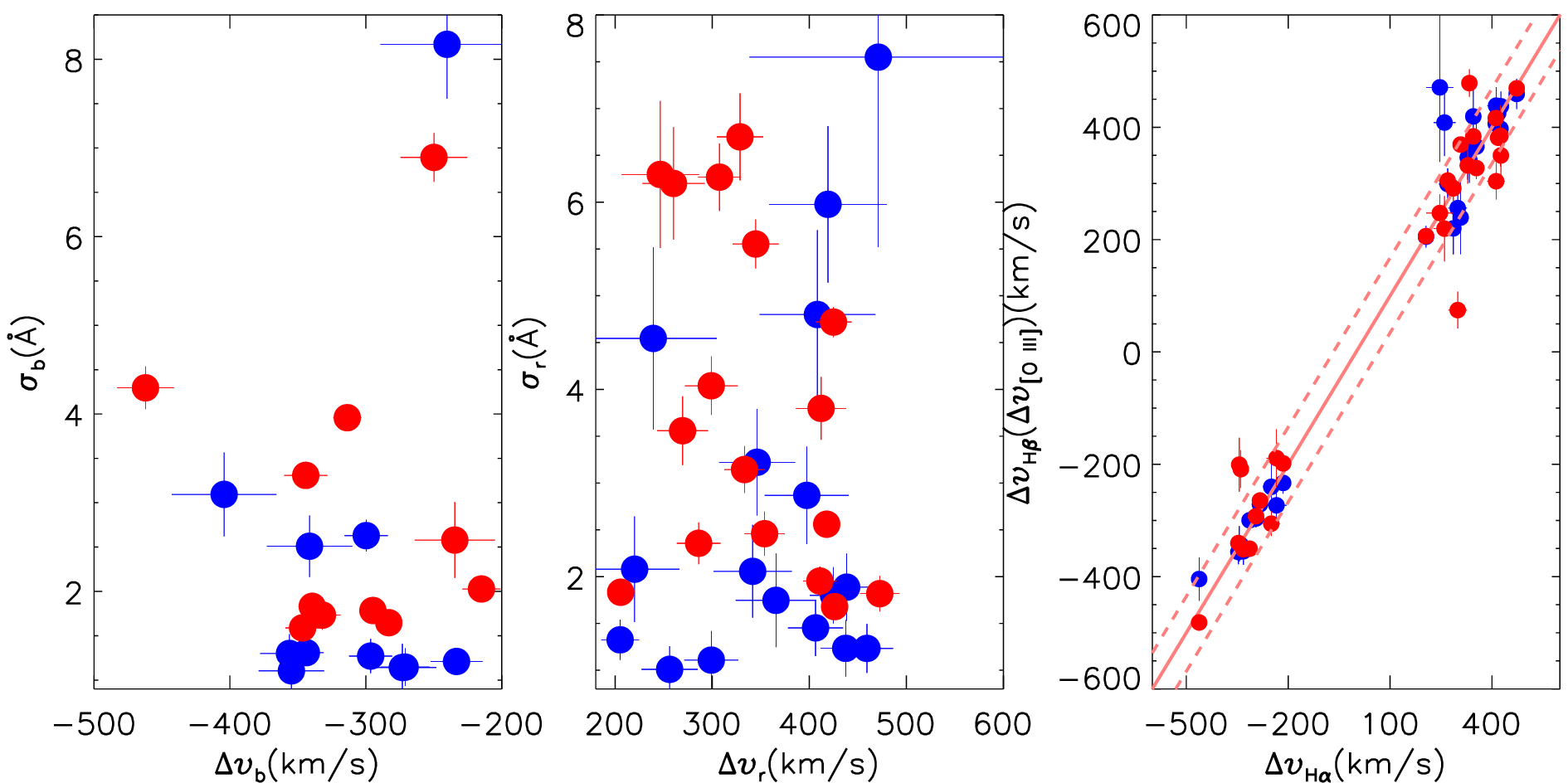}
\caption{The velocity offset plotted against the narrow H$\beta$ skewnesses (top-left panel), the profiles of narrow H$\beta$ emission lines (top-right panel), the relationship between velocity offset ($\Delta\upsilon_{\rm b}$) and line width ($\sigma_{\rm b}$) in blue-shifted systems (bottom-left panel), and the relationship between velocity offset ($\Delta\upsilon_{\rm r}$) and line width ($\sigma_{\rm r}$) in red-shifted systems (bottom-middle panel), the relationships of velocity offsets in narrow emission lines (bottom-right panel). 
In the top-right panel, the Plate-Mjd-Fiberid and H$\beta$ skewnesses are shown in the top-left of corner, the dashed and solid red lines represent 10\% and 20\% of the peak flux, respectively.
In the bottom-left and bottom-middle panels, the solid circles in red mark the results of narrow H$\alpha$ emission lines, while the solid circles in blue correspond to those of narrow H$\beta$ emission lines. In the bottom-right panel, the solid circles in red mark the results of narrow H$\alpha$ and \oiii, and the solid circles in blue correspond to those of narrow H$\alpha$ and H$\beta$ emission lines. The solid pink line indicates the y = x relation, and the dashed pink lines represent 1$\sigma$ scatter.
}
\label{fig6}
\end{figure*}

Furthermore, in the outflow scenario, the gravitational potential naturally produces a positive correlation between velocity offset and emission-line width, and the strong correlation can be found between \oiii~line width and blue-shifted \oiii~in \citet{Ko08}.
The relations of velocity offset and emission-line width in our sample are shown in the bottom-left and bottom-middle panels of Figure \ref{fig6}.
For the blue-shifted emission lines, there are negligible
relationships between the velocity offset and the line width for both narrow H$\beta$ (the Spearman Rank correlation coefficient as -0.15 with null-hypothesis probability of 65.0\%)
and narrow H$\alpha$ emission lines (the Spearman Rank correlation coefficient as 0.07 with null-hypothesis probability of 83.2\%) in our sample.
For the red-shifted emission lines, there is a negligible
relationship between the velocity offset and the line width for narrow H$\beta$ (the Spearman Rank correlation coefficient as 0.17 with null-hypothesis probability of 51.0\%), but a weak negative relationship for narrow H$\alpha$ emission lines (the Spearman Rank correlation coefficient as -0.38 with null-hypothesis probability of 12.4\%) in our sample.
As discussed above, the velocity offset in our sample is not consistent with the outflow scenario.

\subsection{dual core system}
One explanation for our observations of offset narrow H$\beta$ and narrow H$\alpha$ lines is the central core moving with respect to its host galaxy as the result of a merger.
According to the BPT diagram in Figure \ref{fig4}, there are 12 AGNs, 12 composite galaxies, 3 H{\sc ii} galaxies, and 1 unknown-type objects.
It can be accepted that the nucleus activity can be triggered by major mergers of galaxies \citep{Co15,El19}.
Besides, galaxy mergers can initiate starburst activity, whose subsequent stellar evolution provides a potential AGN fueling mechanism. Within tens of Myr, a fraction of the newly formed stars progress to the asymptotic giant branch phase, with their stellar winds potentially enabling efficient mass transfer onto supermassive black holes \citep{Da07}.

\begin{figure*}
\centering\includegraphics[width=15cm,height=16cm]{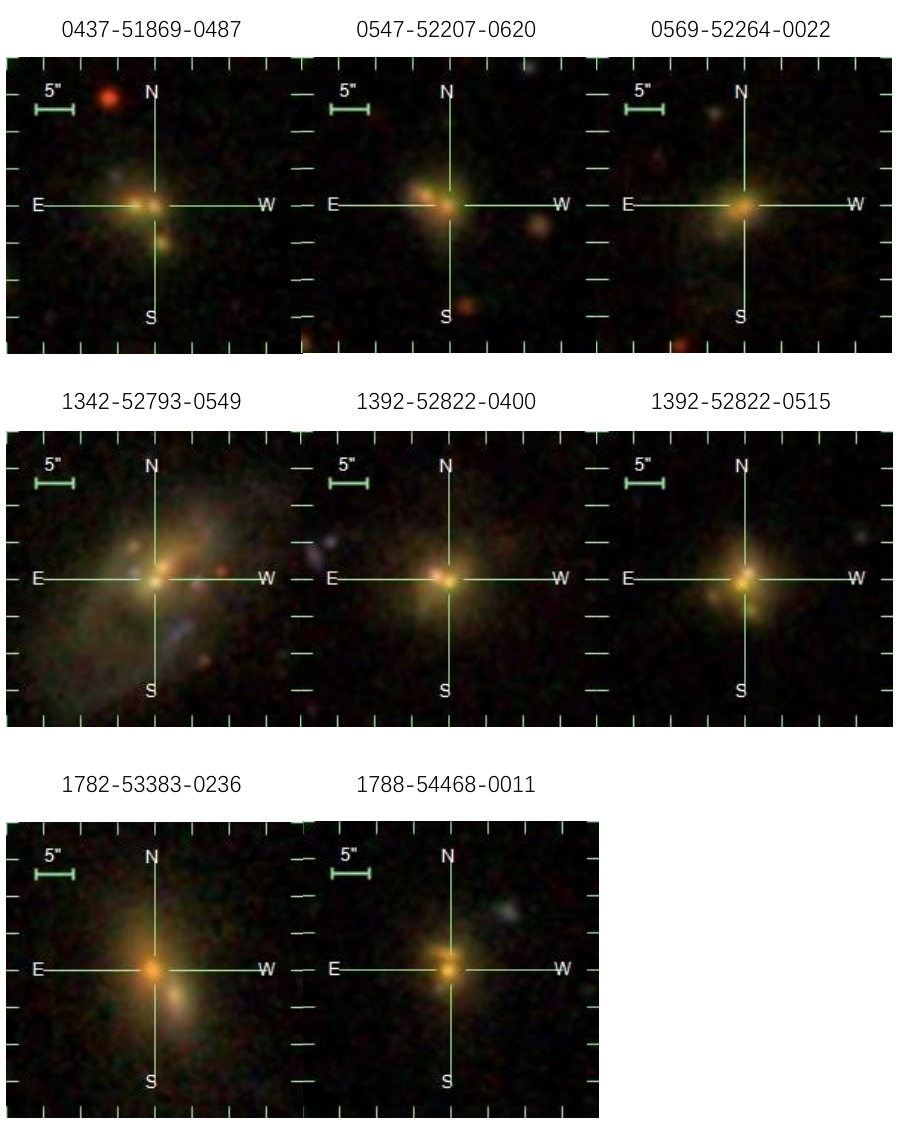}
\caption{Photometric image properties of the eight objects with velocity offset in our sample.}
\label{fig7}
\end{figure*}

In this scenario, only one of the two nuclei is active, while the other nucleus remains inactive.
The less massive black hole may accrete more efficiently and thus exhibit stronger nuclear activity than its more massive counterpart \citep{Ca15}.
Both two nuclei in the merger-remnant galaxy moves with respect to the stellar component of host galaxy.
Consequently, the emission lines from the active core show velocity offsets relative to the absorption system of the host galaxy \citep{Co09,Zh23}, caused by the bulk motion of the active core.
This mechanism produces similar velocity offsets for narrow emission lines with different ionization potentials.
The mean values and distributions of velocity offset of narrow H$\beta$ (mean $\mid$$\Delta\upsilon$$\mid$ as 341.09 $\pm$ 36.21 km/s) and narrow H$\alpha$ (mean $\mid$$\Delta\upsilon$$\mid$ as 329.09 $\pm$ 19.33 km/s) in our sample are similar with the probability of 74.33\% determined by the Student's t-test technique \citep{St08} and with the probability of 71.98\% determined by the Kolmogorov–Smirnov (K-S) test \citep{Ko33,Sm48}, respectively.
The Spearman Rank correlation coefficient of the two parameters is 0.89 with null-hypothesis probability of 2.64$\times10^{-10}$.
Moreover, the mean value and distribution of velocity offset of \oiii~(mean $\mid$$\Delta\upsilon$$\mid$ as 308.96 $\pm$ 22.33 km/s) are similar to that of narrow H$\alpha$ with the probability of 38.20\% determined by the Student's t-test technique and with the probability of 71.98\% determined by the K-S test, respectively. 
And the Spearman Rank correlation coefficient of velocity offset of \oiii~and narrow H$\alpha$ is 0.64 with null-hypothesis probability of 2.26$\times10^{-2}$.
As shown in the bottom-right panel of Figure \ref{fig6},
the strong correlations of velocity offsets between narrow H$\beta$ and narrow H$\alpha$ emission lines, and between \oiii~and narrow H$\alpha$ emission lines strongly support dual core systems, prompting our subsequent visual inspection of images to identify direct morphological signatures of nuclear separations.

It should be noticed that the velocity offsets of two objects (Plate-Mjd-Fiberid: 0742-52263-0540 and 1669-53433-0541) in narrow H$\alpha$ and H$\beta$ are not consistent, but the velocity offsets in \oiii~are consistent with narrow H$\alpha$. It is possible that the H$\beta$ measurement may be affected by a relatively lower signal-to-noise ratio.
Besides, the six objects (Plate-Mjd-Fiberid: 0458-51929-0392, 0538-52029-0623, 0547-52207-0620, 0686-52519-0064, 1685-53463-0515 and 1814-54555-0043) show similar velocity offsets in narrow H$\alpha$ and H$\beta$, but show large deviations in \oiii. A plausible explanation is that the velocity offset may be more strongly affected by outflows. 	
The differences observed in a few individual objects do not affect our overall statistical analysis and the main conclusions of this study.

To test this dual core hypothesis, we conduct systematic visual inspection of SDSS photometric imaging data for all 28 objects in our sample. 
Figure \ref{fig7} presents multi-color composites of eight representative cases exhibiting two compact cores.
In addition, there are two objects (Plate-Mjd-Fiberid: 0305-51613-0168 and 1814-54555-0043) without two cores but a tidal tail, which might be caused by merger.

SDSS J080151.62+431244.4 (Plate-Mjd-Fiberid: 0437-51869-0487), classified as a composite galaxy by the BPT diagram, displays a dual core morphology with a projected separation of about $3^{\prime \prime}$. The NLR of the eastern companion galaxy likely falls within the $3^{\prime \prime}$ diameter SDSS fiber coverage area.

SDSS J082401.64+442510.3 (Plate-Mjd-Fiberid: 0547-52207-0620), classified as a composite galaxy by the BPT diagram, reveals a about $4.8^{\prime \prime}$ separated galaxy pair with merger signatures, and the standard SDSS fiber aperture  is insufficient to encompass both NLRs of two cores.

SDSS J094005.71+031500.2 (Plate-Mjd-Fiberid: 0569-52264-0022), classified as a AGN by the BPT diagram, shows two cores with projected distance about $1.7^{\prime \prime}$, and the SDSS fiber aperture is large enough to encompass the NLR of the companion galaxy.

SDSS J164839.92+300111.8 (Plate-Mjd-Fiberid: 1342-52793-0549), classified as a AGN by the BPT diagram, 
displays a dual core morphology with a projected separation of $2.4^{\prime \prime}$. The northwestern companion galaxy lies sufficiently close to the primary that its NLR likely falls within the SDSS fiber aperture.

SDSS J155708.82+273518.7 (Plate-Mjd-Fiberid: 1392-52822-0400), classified as a composite galaxy by the BPT diagram, reveals two distinct nuclei separated by $2.2^{\prime \prime}$. Given this distance, the SDSS fiber aperture probably captures emission from both the primary galaxy and its northeastern companion \citep{Zh23}.

SDSS J160207.95+272036.2 (Plate-Mjd-Fiberid: 1392-52822-0515), classified as a composite galaxy by the BPT diagram, shows two compact cores with a projected separation of about $1.7^{\prime \prime}$.

SDSS J081914.06+543529.7 (Plate-Mjd-Fiberid: 1782-53383-0236), classified as a H{\sc ii} galaxy by the BPT diagram, shows two cores, and the southwestern companion core is about 
$5.4^{\prime \prime}$ away from the primary core.

SDSS J094741.58+633939.2 (Plate-Mjd-Fiberid: 1788-54468-0011), classified as a AGN by the BPT diagram, exhibits a dual core morphology, where the northern compact core is separated by about $2.5^{\prime \prime}$ in projection.


This observed kinematic signatures of velocity offset can be explained through two primary scenarios within the merger framework: (1) the spectroscopic fiber may have captured only one of the cores due to spatial sampling limitations of the SDSS fiber aperture ($3^{\prime \prime}$ diameter), or (2) asymmetric gas dynamics in the merger system may lead to only one core being currently visible in narrow-line emission.
These eleven objects with single-peaked, velocity offset narrow emission lines provides compelling evidence that the observed kinematics may originate from dual core mergers, though follow-up spectroscopic observations are required for definitive confirmation.
Although 17 objects do not exhibit clear evidence of dual cores or mergers in the SDSS imaging, the possibility of being dual core systems cannot be entirely excluded. Limited spatial resolution may prevent the optical separation of closely spaced cores in some cases.
Combined with our preceding analysis, the observed velocity offset objects likely originate from dual core systems in merging galaxies, making them prime candidates for follow-up high-resolution imaging to confirm nuclear separations.

For the dual core system, it is also possible that two objects can appear spatially proximate along the line of sight, while in reality they reside at significantly different distances \citep{De14}.
The probability of such a close cosmological superposition is, however, very low \citep{He09,De10}. 
The probability increases only if the object lies in a rich cluster environment \citep{He09,Do10}, reaching about $10^{-4.3}$ within $1^{\prime \prime}$ (within a SDSS fiber) \citep{Sh09}. Even in this case, the expected number of such superpositions is very small.
We also test whether our sample is biased toward overdense environments. We count spectroscopic neighbors within $100^{\prime \prime}$ around each object in our sample. For the majority of our systems, fewer than four neighbors are found, and the maximum number is six. These results indicate that the objects in our sample are not locate in rich clusters. Therefore, chance superpositions are unlikely to explain the observed velocity offsets in our sample.

\subsection{future application}
Future high-resolution spectroscopic follow-up can spatially resolve the NLRs and directly test the dual core system hypothesis of our sample by identifying distinct NLRs or velocity components. 
The velocity offset sample identified in this work offers a promising set of candidates for dual core systems and provides an important foundation for investigating merger-driven galaxy evolution. Future research can leverage this sample to conduct systematic comparisons with typical single-core galaxies, aiming to identify differences in multi-wavelength properties such as mid-infrared AGN features, radio morphologies, and X-ray emission. These comparisons may reveal observable signatures uniquely associated with dual core or interacting systems, thereby improving our understanding of the physical conditions that give rise to velocity offsets and dual core systems.


Furthermore, if large velocity offset between narrow emission and absorption lines is confirmed to be an effective diagnostic for identifying dual core systems, this method can be extended to higher-redshift galaxy samples. Applying this selection criterion to deep spectroscopic surveys would enable the identification of dual core system candidates at earlier cosmic epochs, providing crucial constraints on the frequency and impact of mergers over cosmic time. This, in turn, would shed light on the co-evolution of galaxies and their central black holes. Ultimately, this velocity offset approach holds potential for advancing our understanding of galaxy assembly and the mechanisms that drive nuclear activity throughout the universe.

\section{Summary and Conclusions}
We report 28 objects with large velocity offset (200 km/s) of both narrow H$\beta$ and narrow H$\alpha$ emission lines relative to absorption lines through strict criteria at $z<0.3$ in SDSS DR16.
The main summary and conclusions are shown as follows.
\begin{itemize}
\item There is a weak trend between redshift and velocity offset in our sample, possibly due to selection effects: higher-redshift galaxies require higher luminosity and greater mass for detection, with massive mergers showing larger orbital velocities.
\item The BPT diagram classifies our sample into 12 AGN-dominated systems, 12 composite galaxies, 3 H{\sc ii} galaxies and 1 object of ambiguous classification.
\item A strong correlation in our sample is observed between the stellar mass and the velocity dispersion derived from the pPXF method. 
The black hole mass here is derived using the $M_{\rm BH}-M_{\rm \star,bul}$ relation, under the assumption that the total stellar mass is scaled to the bulge mass. Our sample demonstrates consistency with the $M_{\rm BH}-\sigma_{\ast}$ relation.
\item The correlation between $\upsilon_{\rm em}-\upsilon_{\rm abs}$ and the narrow H$\beta$ skewness is found to be negligible in both blue-shifted and red-shifted systems. These results suggest that the rotating disk model may not adequately account for the observed velocity offsets.
\item Our sample displays an asymmetric distribution of velocity offset, with a higher number of red-shifted (17) than blue-shifted (11) emission-line objects. For blue-shifted emission lines, the velocity offset shows negligible correlations with the line width for both narrow H$\beta$ and narrow H$\alpha$. Besides, the velocity offset of red-shifted emission lines exhibits a negligible correlation with the line width of narrow H$\beta$, but a weak negative correlation with that of narrow H$\alpha$. These findings are not consistent with the outflow scenario.
\item Our sample shows similar velocity offset between narrow emission lines, which favors dual core systems. A large velocity offset between emission and absorption lines may serve as an effective indicator of dual core systems.
\item After visually checking the SDSS multi-color images, we find eight objects with two cores and two objects with signs of merging. 
\item The emission-line characteristics in our sample identify these objects as potentially robust dual core system candidates exhibiting kinematic offsets. Applying this velocity offset features to higher-redshift systems may enable earlier identification of merging nuclei during cosmic epochs when galaxy interactions were more frequent.
\end{itemize}

\section*{Acknowledgements}
Zheng, Zhang and Yuan gratefully acknowledge the anonymous referee for reading our paper carefully and patiently, and giving us constructive and valuable comments and suggestions to greatly improve our paper.
This work is supported by the National Natural Science Foundation of China (Nos. 12273013, 12173020, 12373014). We have made use of the data from SDSS DR16.
The SDSS DR16 web site is (http://skyserver.sdss.org/dr16/en/home.aspx), and the SQL Search tool can be found in (\url{http://skyserver.sdss.org/dr16/en/tools/search/sql.aspx}).
The MPFIT website is (\url{http://cow.physics.wisc.edu/~craigm/idl/idl.html}).

\section*{Data Availability}
The data underlying this article will be shared on reasonable request to the corresponding author (aexueguang@qq.com).

\end{document}